\documentclass[review]{elsarticle}

\usepackage{lineno,hyperref}
\modulolinenumbers[5]
\usepackage{breqn}
\usepackage{tabularx}
\usepackage{url}
\usepackage{siunitx}
\usepackage{color}
\usepackage{bm}
\biboptions{sort&compress}
\usepackage[version=4]{mhchem}
\usepackage[subrefformat=parens]{subcaption}
\newcolumntype{Y}{&gt;{\centering\arraybackslash}X}
\journal{Journal of \LaTeX\ Templates}
\begin{document}
	
	\begin{frontmatter}
		
		\title{FGM modeling considering preferential diffusion, flame stretch, and non-adiabatic effects for hydrogen-air premixed flame wall flashback}
		
		\author{Kazuhiro Kinuta\fnref{adr}\corref{firstcorr}}
		\author{Reo Kai\fnref{adr2}}
		\author{Kotaro Yada\fnref{adr}}
		\author{Ryoichi Kurose\fnref{adr}}
		\address[adr]{Department of Mechanical Engineering and Science, Kyoto University, Kyoto daigaku-Katsura, Nishikyo-ku, Kyoto 615–8540, Japan}
		\address[adr2]{Department of Advanced Environmental Science and Engineering, Kyushu University, Kasuga-koen, Kasuga, Fukuoka 816-8580, Japan}
		\cortext[firstcorr]{E-mail address: kinuta.kazuhiro.36w@st.kyoto-u.ac.jp (K. Kinuta)}
		
		\begin{abstract}
			Preferential diffusion plays an important role especially in hydrogen flames.
			Flame stretch significantly affects the flame structure and induces preferential diffusion.
			A problematic phenomenon occurring in real combustion devices is flashback, which is influenced by non-adiabatic effects, such as wall heat loss.
			In this paper, an extended flamelet-generated manifold (FGM) method that explicitly considers the preferential diffusion, flame stretch, and non-adiabatic effects is proposed.
			In this method, the diffusion terms in the transport equations of scalars, viz. the progress variable, mixture fraction, and enthalpy, are formulated employing non-unity Lewis numbers that are variable in space and different for each chemical species.
			The applicability of the extended FGM method to hydrogen flames is investigated using two- and three-dimensional numerical simulations of hydrogen-air flame flashback in channel flows.
			The results of the extended FGM method are compared with those of detailed calculations and other FGM methods.
			The two-dimensional numerical simulations show that considering both preferential diffusion and flame stretch improves the prediction accuracy of the mixture fraction distribution and flashback speed.
			The three-dimensional numerical simulations show that the prediction accuracy of the flashback speed, backflow region, and distributions of physical quantities near the flame front is improved by employing the extended FGM method, compared with the FGM method that considers only the heat loss effect.
			In particular, the extended FGM method successfully reproduced the relationship between the reaction rate and curvature.
			These results demonstrate the effectiveness of the extended FGM method.
			
		\end{abstract}
		
		
	\end{frontmatter}
	
	\clearpage
	
	\section{Introduction}
		To achieve carbon neutrality, hydrogen (\ce{H2})-fueled combustion devices, particularly combustors, are being developed rapidly. 
		From the perspecrive of NOx emissions, fuel-lean premixed combustion is preferable. 
		However, the high flame speed of \ce{H2} flame increases the risk of flashback.
		Therefore, predicting and preventing flashback are mandatory for developing stable \ce{H2}-fueled combustors. 
		As the occurrence of flashback results in the destruction of the combustor, numerical simulation can be an effective tool to predict the occurrence of flashback in addition to experimental tests, if it has high accuracy, e.g., \cite{gruber2012direct,kitano2015effect,ahmed2019statistical}. 
		
		Although direct numerical simulation (DNS) reproduces the combustion field with high accuracy, its computational cost is extremely high. 
		Therefore, a combustion model is required to simulate large-scale combustors.
		In numerical simulations of \ce{H2} flames, the employed combustion model must be able to reproduce the preferential diffusion (PD) effect because of the high diffusivity of \ce{H2}.
		A flamelet-generated manifold (FGM) method \cite{Oijen2000} that considers PD (FGM-P method) has been proposed and applied to pure \ce{H2} or \ce{H2}-blended fuel combustion \cite{Bastiaans2007,Swart2010,bastiaans2012numerical,abtahizadeh2015development,mukundakumar2021new,almutairi2023modelling,kai2023flamelet}.
		In addition to the PD effect, the flame stretch (FS) and non-adiabatic (NA) effects need to be considered in the FGM method in the numerical simulation of \ce{H2} boundary layer flashback	because the flame front is curved or distorted, and the heat loss through the wall suppresses the reaction.
		
		Bastiaans et al. \cite{Bastiaans2007,bastiaans2012numerical} and de Swart et al. \cite{Swart2010} reproduced the developing cellular structure of wrinkled flame fronts under adiabatic conditions using the FGM-P method and considering the FS effect.
		Kai et al. \cite{kai2024effects} demonstrated the importance of considering the PD and FS effects in the FGM method by performing two-dimensional numerical simulations of expanding flames.
		Mukundakumar et al. \cite{mukundakumar2021new} considered the NA effect in the FGM-P method (FGM-PN method) and reproduced a two-dimensional slot-burner flame stabilized on an isothermal wall. 
		Although the considerations of the FS and NA effects in the FGM-P method have been validated separately, both effects have never been considered together, and the applicability of the FGM-P method, which considers the FS and NA effects (FGM-PFN method), to the numerical simulation of lean \ce{H2} flame flashback has never been investigated. 
		
		This study aims to establish an FGM approach applicable to the boundary layer flashback of a lean \ce{H2}-air premixed flame.
		To this end, two- and three-dimensional numerical simulations are performed under the conditions of an equivalence ratio of 0.5, an unburnt gas temperature of \SI{750}{K}, and a pressure of \SI{1}{atm}.
		Two-dimensional numerical simulations are performed by employing different types of FGM methods and by directly calculating a detailed reaction mechanism (detailed calculation) under laminar conditions, and the validity of each FGM method is compared. 
		A three-dimensional numerical simulation employing the FGM-PFN method is conducted at a friction Reynolds number of 180, and its applicability to turbulent boundary layer flashback is investigated.

	\section{Numerical methods}	\label{sec_numeric}
	This study utilizes detailed calculation and three different FGM methods, in which fine grids are set to satisfy the DNS requirements in terms of turbulence.
	
	\subsection{Detailed calculation\label{subsec:detailed}} \addvspace{10pt}
	In the detailed calculation, in which no combustion model is used, the following transoirt equations of mass, momentum, energy, and mass of chemical species are solved. 
	\begin{equation}
		\label{eq:mass_detail}
		\frac{\partial \rho}{\partial t} + \nabla \cdot \left( \rho \bm{u} \right) = 0,
	\end{equation} 
	\begin{equation}
		\label{eq:momentum_detail}
		\frac{\partial \rho \bm{u}}{\partial t} + \nabla \cdot \left( \rho \bm{u u} \right) = - \nabla p + \nabla \cdot \bm{\tau},
	\end{equation} 
	\begin{align*}
		\label{eq:h_detail}
		\frac{\partial \rho h}{\partial t} &+ \nabla \cdot \left( \rho h \bm{u} \right) = 	\frac{\partial p}{\partial t} + \bm{u} \cdot \nabla p \\
		&+ \nabla \cdot \left[ \rho D_h \left( \nabla h - \sum_{k}h_k\nabla Y_k \right) - 	\rho \sum_{k}h_k Y_k \bm{V}_k \right] + \bm{\tau} \colon \nabla \bm{u},
		\stepcounter{equation}\tag{\theequation} 
	\end{align*}
	\begin{equation}
		\label{eq:Yk_detail}
		\frac{\partial \rho Y_k}{\partial t} + \nabla \cdot \left( \rho Y_k \bm{u} 	\right) = - \nabla \cdot \left(\rho Y_k \bm{V}_k \right) + \rho \dot{\omega}_k,
	\end{equation} 
	where $\rho$ is the density, $\bm{u}$ is the velocity, $p$ is the pressure, $\bm{\tau}$ is the viscous stress, $h$ is the specific enthalpy, $h_k$ is the specific enthalpy of species $k$, $Y_k$ is the mass fraction of species $k$, $\dot{\omega}_k$ is the reaction rate of species $k$, and $D_h$ is the thermal diffusivity.
	The diffusion velocity of species $k$, $\bm{V}_k$, is calculated using the Maxwell--Stefan equation neglecting the species diffusion caused by the temperature and pressure gradients as follows: 
	\begin{equation}
		\label{eq:MS}
		\nabla X_k = \sum_{j \ne k}{\frac{X_j X_k}{D_{jk}} \left( \bm{V}_j - \bm{V}_k \right)}.
	\end{equation}
	Here, $X_k$ is the mole fraction of species $k$, and $D_{jk}$ is the binary diffusion coefficient of species $k$ into $j$.
	The \ce{H2} reaction is represented using Conaire's mechanism \cite{o2004comprehensive}, which includes 9 species and 19 reactions.

	\subsection{FGM-PFN method \label{subsec:FGM-PFN}} \addvspace{10pt}
	
	The FGM-PFN method proposed in this paper explicitly considers the preferential diffusion (PD), flame stretch (FS), and non-adiabatic (NA) effects.
	In the FGM-PFN method, the flame properties are regarded as functions of three control variables: $C$, $Z$, and $\Delta h$. 
	$C$ is the progress variable defined as $C = Y_{\ce{H2O}}$, where $Y_{\ce{H2O}}$ is the mass fraction of \ce{H2O}. 
	$Z$ is the mixture fraction based on Bilger's definition \cite{bilger1989structure}.
	$\Delta h$ is the difference in enthalpy written as $\Delta h = h_0 - h$, where $h_0$ is the specific enthalpy without heat loss and $h$ is the specific enthalpy obtained by solving the transport equation.
	To generate the flamelet library, one-dimensional numerical simulations of premixed flames are performed under various flame stretch rates and degrees of heat loss using the FlameMaster code \cite{flamemaster}.
	To consider the heat loss, the same method as M3 (third method) in \cite{proch2015modeling} is employed. 
	The Conaire mechanism \cite{o2004comprehensive} is used in these calculations.
	The results are tabulated in the $C$-$Z$-$\Delta h$ space.

	In the FGM-PFN method, in addition to the conservation equations of mass and momentum, the transport equations of $C$, $Z$, and $h$, which consider the PD effect, are solved. The transport equation of $C$ is expressed as
	\begin{equation}
		\label{eq:C_FGM-PFN}
		\frac{\partial \rho C}{\partial t} + \nabla \cdot \left( \rho C \bm{u} \right) = \nabla \cdot \left( \frac{\rho D_C}{W} \nabla \left( C W \right) \right) + \rho \dot{\omega}_C,
	\end{equation}
	where $W$ denotes the average molar mass of the gas mixture.
	Here, $D_C$ is the diffusion coefficient of $C$, which is expressed as $D_C$ = $D_h$/$Le_C$ with a spatially variable non-unity Lewis number.
	$\dot{\omega}_C$ is the reaction rate of $C$, which is equal to the reaction rate of \ce{H2O} in this study.
	The equations for $Z$ and $h$ are derived through the discussion in this section.
	The mixture fraction based on Bilger's definition \cite{bilger1989structure} can be expressed as a linear combination of the mass fractions of the chemical species:
	\begin{equation} \label{eq:Bilger_Z}
		Z = \sum_{k}{z_k Y_k} +\zeta,
	\end{equation}
	where $z_k$ and $\zeta$ are coefficients.	
	Using Eqs.~(\ref{eq:Yk_detail}) and (\ref{eq:Bilger_Z}), the transport equation of $Z$ can be written as
	\begin{equation} \label{eq:Z1_FGM-PFN}
		\frac{\partial \rho Z}{\partial t} + \nabla \cdot \left( \rho Z \bm{u} 	\right) = - \nabla \cdot \left(\rho \sum_{k}{z_k Y_k \bm{V}_k} \right).
	\end{equation}
	The diffusion velocity of the chemical species $\bm{V}_k$ can be calculated using the Hirschfelder--Curtiss approximation written as:
	\begin{equation} \label{eq:HC}
		X_k \bm{V}_k = - D_k \nabla X_k,
	\end{equation}
	\begin{equation} \label{eq:Dk}
		D_k = \frac{1 - Y_k}{\sum_{j \ne k}{X_j / D_{jk}}},
	\end{equation}
	where $D_k$ is the mixture-averaged diffusion coefficient of species $k$ and represents the diffusivity of species $k$ into the gas mixture.
	The Lewis numbers of species $k$ are evaluated using $D_k$ as follows:
	\begin{equation} \label{eq:lewis}
		Le_k = \frac{D_h}{D_k} = \frac{\lambda}{\rho c_p D_k},
	\end{equation}
	where $\lambda$ is the thermal conductivity, and $c_p$ is the isobaric specific heat capacity. 
	In addition, $X_k$ and $Y_k$ have the following relationship.
	\begin{equation} \label{eq:Yk_Xk}
		Y_k = \frac{W_k}{W} X_k,
	\end{equation}
	where $W_k$ is the molar mass of species $k$.
	Using Eqs.~(\ref{eq:HC}), (\ref{eq:lewis}), and (\ref{eq:Yk_Xk}), the transport equations of $Z$ and $h$, i.e., Eqs. (\ref{eq:Z1_FGM-PFN}) and (\ref{eq:h_detail}), can be rewritten as follows:
	\begin{equation} \label{eq:Z2_FGM-PFN}
		\frac{\partial \rho Z}{\partial t} + \nabla \cdot \left( \rho Z \bm{u} 	\right) = \nabla \cdot \left(\frac{\lambda}{c_p W} \sum_{k}{\frac{z_k}{Le_k} \nabla \left( Y_k W \right)} \right),
	\end{equation}
	\begin{align*} \label{eq:h2_FGM-PFN}
		\frac{\partial \rho h}{\partial t} &+ \nabla \cdot \left( \rho h \bm{u} \right) = \\
		&\nabla \cdot \left[ \rho D_h \left( \nabla h - \sum_{k}h_k\nabla Y_k \right) + \frac{\lambda}{c_p W} \sum_{k}{ \frac{h_k}{Le_k} \nabla \left( Y_k W \right)} \right] + \bm{\tau} \colon \nabla \bm{u}.
		\stepcounter{equation}\tag{\theequation}
	\end{align*}
	In addition to other studies on the flamelet approach, the differentiation terms of $p$ are neglected in Eq.~(\ref{eq:h2_FGM-PFN}).
	In the FGM-PFN method, the properties can be expressed as functions of control variables $C$, $Z$, and $\Delta h$.
	Therefore, the following equation holds using the chain rule. 
	\begin{align} \label{eq:chain_Yk}
		\notag	\nabla \varphi 
		&= \frac{\partial \varphi}{\partial C} \nabla C + \frac{\partial \varphi}{\partial Z} \nabla Z +\frac{\partial \varphi}{\partial \left( \Delta h \right)} \nabla \left( \Delta h \right) \\ 
		&= \frac{\partial \varphi}{\partial C} \nabla C + \frac{\partial \varphi}{\partial Z} \nabla Z +\frac{\partial \varphi}{\partial \left( \Delta h \right)} \nabla \left( h_0 \left(C, Z\right) - h \right) \\ \notag
		&= \left(\frac{\partial \varphi}{\partial C} + \frac{\partial \varphi}{\partial \left(\Delta h\right)} \frac{\partial h_0}{\partial C} \right) \nabla C
		+ \left(\frac{\partial \varphi}{\partial Z} + \frac{\partial \varphi}{\partial \left(\Delta h\right)} \frac{\partial h_0}{\partial Z} \right) \nabla Z
		-\frac{\partial \varphi}{\partial \left( \Delta h \right)} \nabla h,
	\end{align}
	where $\varphi$ is $Y_k$ or $Y_k W$.
	Substituting Eq.~(\ref{eq:chain_Yk}) into Eq.~(\ref{eq:Z2_FGM-PFN}), we obtain the following transport equation of $Z$, considering the PD effect:
	\begin{align} \label{eq:Z_FGM-PFN}
		\frac{\partial \rho Z}{\partial t} + \nabla \cdot \left( \rho Z \bm{u} \right) 
		= \nabla \cdot \left( d_{ZC} \nabla C + d_{ZZ} \nabla Z + d_{Zh} \nabla h \right),
	\end{align}
	\begin{align} \label{eq:dzc}
		d_{ZC} =  \frac{\lambda}{c_p W} \sum_{k} \left[ \frac{z_k}{Le_{k}} \left( \frac{\partial Y_k W}{\partial C}  +\frac{\partial Y_k W}{\partial \left( \Delta h \right)}  \frac{\partial h_0}{\partial C} \right) \right],
	\end{align}
	\begin{align} \label{eq:dzz}
		d_{ZZ} = \frac{\lambda}{c_p W} \sum_{k} \left[ \frac{z_k}{Le_{k}} \left( \frac{\partial Y_k W}{\partial Z} +\frac{\partial Y_k W}{\partial \left( \Delta h \right)}  \frac{\partial h_0}{\partial Z} \right) \right],
	\end{align}
	\begin{align} \label{eq:dzh}
		d_{Zh} = \frac{\lambda}{c_p W} \sum_{k} \left[ - \frac{z_k}{Le_{k}} \frac{\partial Y_k W}{\partial \left( \Delta h \right)}  \right].
	\end{align}
	Substituting Eq.~(\ref{eq:chain_Yk}) into Eq.~(\ref{eq:h2_FGM-PFN}), the transport equation of $h$, considering the PD effect, can be written as
	\begin{align} \label{eq:h_FGM-PFN}
		\frac{\partial \rho h}{\partial t} + \nabla \cdot \left( \rho h \bm{u} \right) = 
		\nabla \cdot \left( d_{hC} \nabla C + d_{hZ} \nabla Z + d_{hh} \nabla h \right) + \bm{\tau} \colon \nabla \bm{u},
	\end{align}
	\begin{align} \label{eq:dhc}
		\notag
		d_{hC} = - \frac{\lambda}{c_p} \sum_{k} \left[ h_k \left( \frac{\partial Y_k}{\partial C} +\frac{\partial Y_k}{\partial \left( \Delta h \right)} \frac{\partial h_0}{\partial C} \right) \right] \\ 
		+ \frac{\lambda}{c_p W} \sum_{k} \left[ \frac{h_k}{Le_k} \left( \frac{\partial Y_k W}{\partial C} +\frac{\partial Y_k W}{\partial \left( \Delta h \right)} \frac{\partial h_0}{\partial C} \right) \right],
	\end{align}
	\begin{align} \label{eq:dhz}
		\notag
		d_{hZ} = - \frac{\lambda}{c_p} \sum_{k} \left[ h_k \left( \frac{\partial Y_k}{\partial Z} +\frac{\partial Y_k}{\partial \left( \Delta h \right)} \frac{\partial h_0}{\partial Z} \right) \right] \\ 
		+ \frac{\lambda}{c_p W} \sum_{k} \left[ \frac{h_k}{Le_k} \left( \frac{\partial Y_k W}{\partial Z}+\frac{\partial Y_k W}{\partial \left( \Delta h \right)} \frac{\partial h_0}{\partial Z} \right) \right],
	\end{align}
	\begin{align} \label{eq:dhh}
		d_{hh} = \frac{\lambda}{c_p} + \frac{\lambda}{c_p} \sum_{k} \left[ h_k \frac{\partial Y_k}{\partial \left( \Delta h \right)}   \right]  
		- \frac{\lambda}{c_p W} \sum_{k} \left[ \frac{h_k}{Le_k} \frac{\partial Y_k W}{\partial \left( \Delta h \right)}  \right].
	\end{align}
	The coefficients $d_{ZC}$, $d_{ZZ}$, $d_{Zh}$, $d_{hC}$, $d_{hZ}$, and $d_{hh}$ are tabulated in advance in the flamelet database.
	They are obtained from the database when calculating the transport equations.
	The Lewis numbers $Le_k$ are evaluated using mixture-averaged diffusion, whereas constant $Le_k$ are used in most previous studies. 
	Therefore, the spatial variation in $Le_k$ is considered in this study.
	The importance of the variable $Le_k$ is demonstrated in a previous study \cite{kinuta2024}.
	Moreover, the $W$ gradient is not neglected in this study.
	
	\subsection{FGM-PN method\label{subsec:FGM-PN}} \addvspace{10pt}
	
	The FGM-PN method considers the PD and NA effects.
	In contrast to the FGM-PFN method, one-dimensional numerical simulations of unstretched flames with various degrees of heat loss are performed to generate a flamelet library.
	The results are tabulated in the $C$-$\Delta h$ space.
	
	In the FGM-PN method, the transport equations of $C$ and $h$ are solved in addition to the conservation equations of mass and momentum.
	The transport equation of $C$ is the same as that for the FGM-PFN method, whereas the transport equation of $h$ is written as follows:
	\begin{align}
		\label{eq:h_FGM-PN}
		\frac{\partial \rho h}{\partial t} + \nabla \cdot \left( \rho h \bm{u} \right) = \nabla \cdot \left( d_{hC} \nabla C + d_{hh} \nabla h \right)
		 + \bm{\tau} \colon \nabla \bm{u}.
	\end{align}
	
	\subsection{FGM-N method\label{subsec:FGM-N}} \addvspace{10pt}
	
	The FGM-N method considers the NA effect but not the PD and FS effects.
	Similar to the FGM-PN method, the transport equations of $C$ and $h$, i.e., Eqs. (\ref{eq:C_FGM-PFN}) and (\ref{eq:h_FGM-PN}), are solved.
	The only difference from the FGM-PN method is the application of the unity Lewis numbers assumption.
	To generate the flamelet library, one-dimensional numerical simulations of unstretched flames with various degrees of heat loss are performed using unity Lewis numbers.
	Moreover, $D_C$ in Eq.~(\ref{eq:C_FGM-PFN}) is modified as $D_C = D_h$, and $Le_k$ in Eqs.~(\ref{eq:dhc}) and (\ref{eq:dhh}) are assumed to be equal to 1.
	
	\subsection{Numerical setups\label{subsec:setup}} 
\begin{figure}[h]
		\centering
		\includegraphics[width=1.0\linewidth]{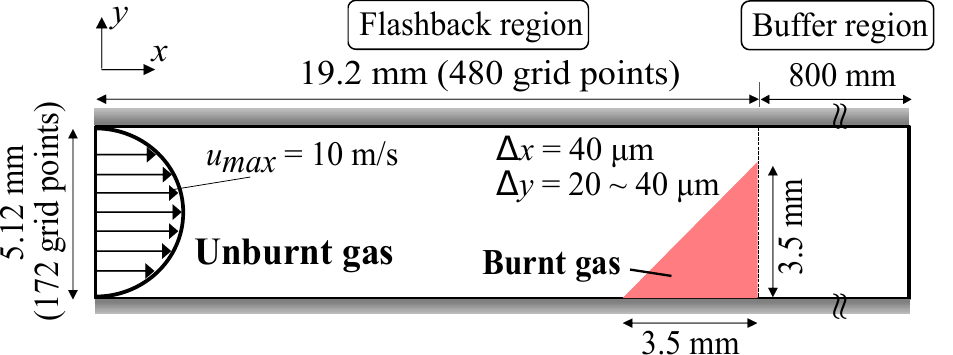}
		\caption{Schematic of computational domain and conditions of a two-dimensional \ce{H2}-air flame flashback in laminar channel flow.}
		\label{fig:2Df_domain}
	\end{figure}

\addvspace{10pt}
	Figure \ref{fig:2Df_domain} shows the computational domain of the two-dimensional numerical simulations of an \ce{H2}-air premixed flame flashback in laminar flow.
	The computational domain consists of a flashback region, and a buffer region, and is discretized on a non-uniform staggered grid with a Cartesian coordinate system.
	The flashback region measures \SI{19.2}{mm} in the $x$-direction and \SI{5.12}{mm} in the $y$-direction.
	The grid spacing of the flashback region is \SI{40}{\micro m} in the $x$-direction and 20--\SI{40}{\micro m} from the wall to the center of the channel in the $y$-direction.
	A laminar flow of unburnt gas enters the flashback region from the left boundary.
	
\begin{figure}[h]
		\centering
		\includegraphics[width=1.0\linewidth]{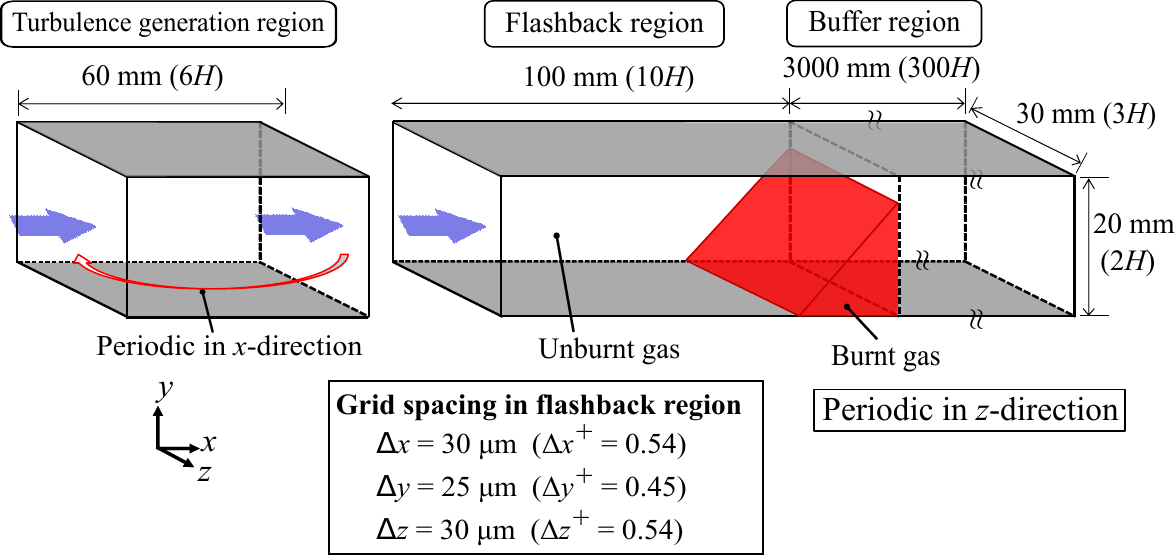}
		\caption{Schematic of computational domain and conditions of a three-dimensional \ce{H2}-air flame flashback in turbulent channel flow.}
		\label{fig:3Df_domain}
	\end{figure}

	Figure \ref{fig:3Df_domain} shows the computational domain of the three-dimensional numerical simulations of an \ce{H2}-air premixed flame flashback in turbulent flow.
	The computational domain consists of a turbulence generation region, flashback region, and buffer region, and is discretized on a non-uniform staggered grid with a Cartesian coordinate system.
	The grid spacing of the turbulence generation region is \SI{500}{\micro m} in the $x$-direction ($\Delta x^+$ = 9), \SI{25}{\micro m} in the $y$-direction ($\Delta y^+$ = 0.45), and \SI{30}{\micro m} in the $z$-direction ($\Delta z^+$ = 0.54), where the superscript + is the wall unit length.
	For the flashback region, the grid spacing is \SI{25}{\micro m} in the $y$-direction ($\Delta y^+$ = 0.45), and \SI{30}{\micro m} in the $x$- and $z$-directions ($\Delta x^+$ = $\Delta z^+$ = 0.54).
	The grid spacing satisfies the recommendation by Moser et al.~\cite{moser1999direct} that $y^+ = u_{\tau} y/\nu_{w}$ should be satisfied near the wall to ensure adequate resolution of the boundary layer.
	Here, $u_{\tau}$ is the friction velocity, and $\nu_{w}$ is the kinematic viscosity at the wall.
	In the turbulence generation region, wall-bounded turbulence at a friction Reynolds number of 180 is developed.
	The turbulent flow of the unburnt gas obtained from a calculation in the turbulence generation region flows into the flashback region with the dimensions of \SI{100}{mm} $\times$ \SI{30}{mm} $\times$ \SI{20}{mm}.
	The total number of grid points for the three-dimensional domain is approximately 3.2 billion.
	
	Both the two- and three-dimensional numerical simulations are performed under an ambient pressure of 1 atm, an unburnt gas temperature of 750 K, and an equivalence ratio of 0.5.
	The no-slip isothermal walls at 750 K are set at the boundaries in the $y$-direction.
	Buffer regions are placed on the right side of the computational domains to reduce the effects of the outlet boundary.
	All the numerical simulations are performed using the in-house code $\rm{FK^3}$~\cite{Kurose_FK3}.
	This code has been used to investigate the flame-wall interactions, including flashback behavior~\cite{kitano2015effect, ahmed2019statistical, ahmed2020surface, pillai2022numerical, kai2022analysis, ahmed2023flame}.
	The code is based on a semi-implicit solver for compressible flows that employs the fractional-step method~\cite{moureau2007efficient}.
	The spatial derivative of the convective term in the momentum equation is evaluated using a fourth-order central difference scheme.
	The WENO scheme~\cite{jiang1996efficient} is applied to the convective terms in the governing equations of the scalar quantities.
	A lower-order scheme is used for the spatial derivative near the wall.
	The time integration of these convective terms is conducted using the third-order TVD Runge--Kutta scheme~\cite{gottlieb1998total}.
	For a detailed calculation, the VODE solver \cite{brown1989vode} is employed for the time integration of the source terms owing to chemical reactions.
	The time step size is $\Delta t$ = 1 $\times 10^{-7}$ s for the two-dimensional calculations and $\Delta t$ = 5 $\times 10^{-8}$ s for the three-dimensional calculations.
	
	Two-dimensional numerical simulations are performed using four different methods: a detailed calculation, and the FGM-PFN, FGM-PN, and FGM-N methods.
	Three-dimensional numerical simulations are performed using three different methods: a detailed calculation, and the FGM-PFN and FGM-N methods.
	Note that all the FGM methods consider the NA effect (i.e., heat loss) because it is necessary for the FGM approach to properly predict wall flashback \cite{fukuba2024prediction}.

	The CPU times required for the two-dimensional numerical simulations using the detailed calculation, and the FGM-PFN, FGM-PN, and FGM-N methods are 245.0 h (4.08 h of real time), 210.0 h (3.50 h of real time), 200.0 h (3.33 h of real time), and 199.3 h (3.32 h of real time), respectively, by parallel computation using 60 cores on the supercomputer system Camphor3 at Kyoto University.
	The CPU times required for the three-dimensional numerical simulations using the detailed calculation, and the FGM-PFN, and FGM-N methods are 7.17 Mh (280 h of real time), 1.55 Mh (61 h of real time), and 1.36 Mh (53 h of real time), respectively, by parallel computation using 25,600 cores on the supercomputer Fugaku at RIKEN, Japan.
	The three-dimensional simulations run for \SI{1.5}{ms} of physical time.
	The FGM-PFN method requires approximately 85 \% computational time in the two-dimensional numerical simulations and 20 \% computational time in the three-dimensional numerical simulations. 
	This suggests that the FGM-PFN method is more effective in large scale numerical simulations.

	\section{Results and discussion}	\label{sec_result}
		
	\subsection{Two-dimensional turbulent boundary layer flashback} \label{subsec_2D}
\begin{figure}[htbp]
			\centering
			\begin{tabular}{cccc}
				\centering
				\begin{minipage}[t]{0.5\hsize}
					\centering
					\includegraphics[width=1.0\linewidth]{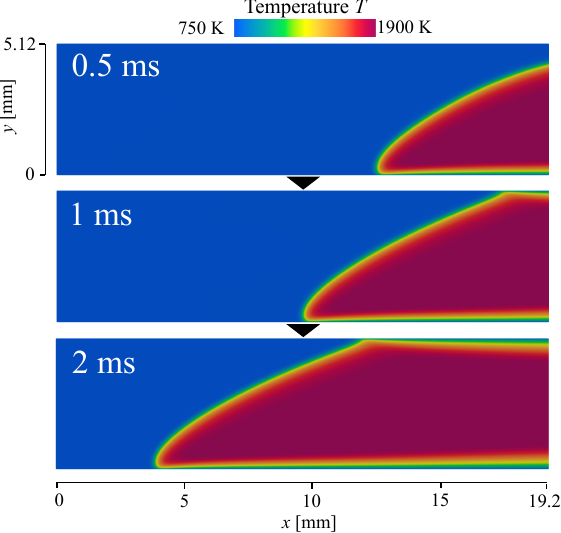}
					\subcaption{Detailed}
				\end{minipage} & 
				\begin{minipage}[t]{0.5\hsize}
					\centering
					\includegraphics[width=1.0\linewidth]{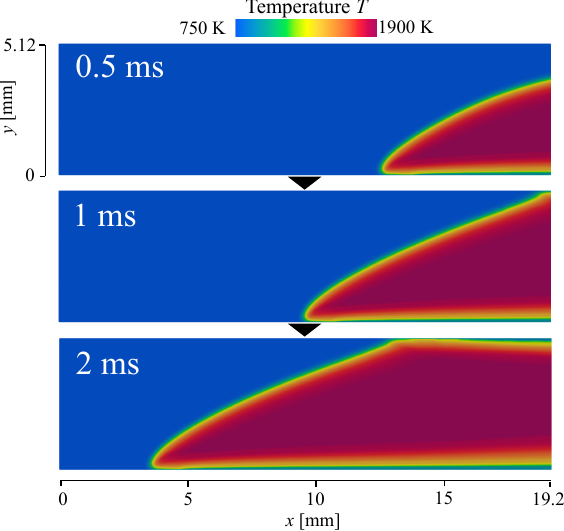}
					\subcaption{FGM-PFN}
				\end{minipage} & \\
				\begin{minipage}[t]{0.5\hsize}
					\centering
					\vspace{3 pt}
					\includegraphics[width=1.0\linewidth]{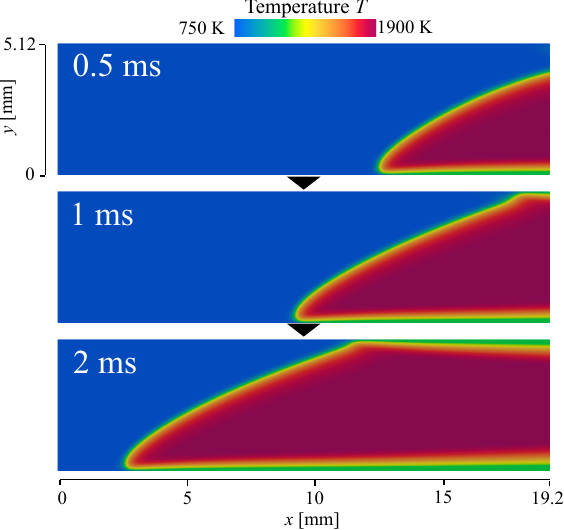}
					\subcaption{FGM-PN}
				\end{minipage} & 
				\begin{minipage}[t]{0.5\hsize}
					\centering
					\vspace{3 pt}
					\includegraphics[width=1.0\linewidth]{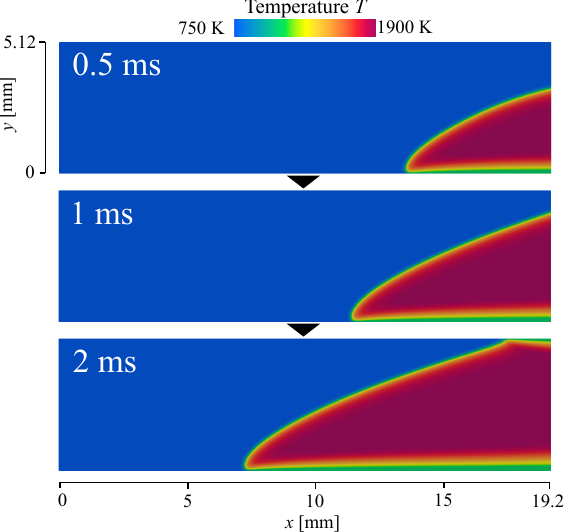}
					\subcaption{FGM-N}
				\end{minipage} \\
			\end{tabular}\\
			\caption{Sequential images of temperature $T$ distributions obtained from (a) detailed calculation, (b) FGM-PFN, (c) FGM-PN, (d) FGM-N methods at $t$ = \SI{0.5}{ms}, \SI{1}{ms}, and \SI{2}{ms}.}
			\label{fig:T_field}
		\end{figure}

\begin{figure}[htbp]
			\centering
			\begin{tabular}{cc}
				\centering
				\begin{minipage}[t]{0.7\hsize}
					\centering
					\includegraphics[width=1.0\linewidth]{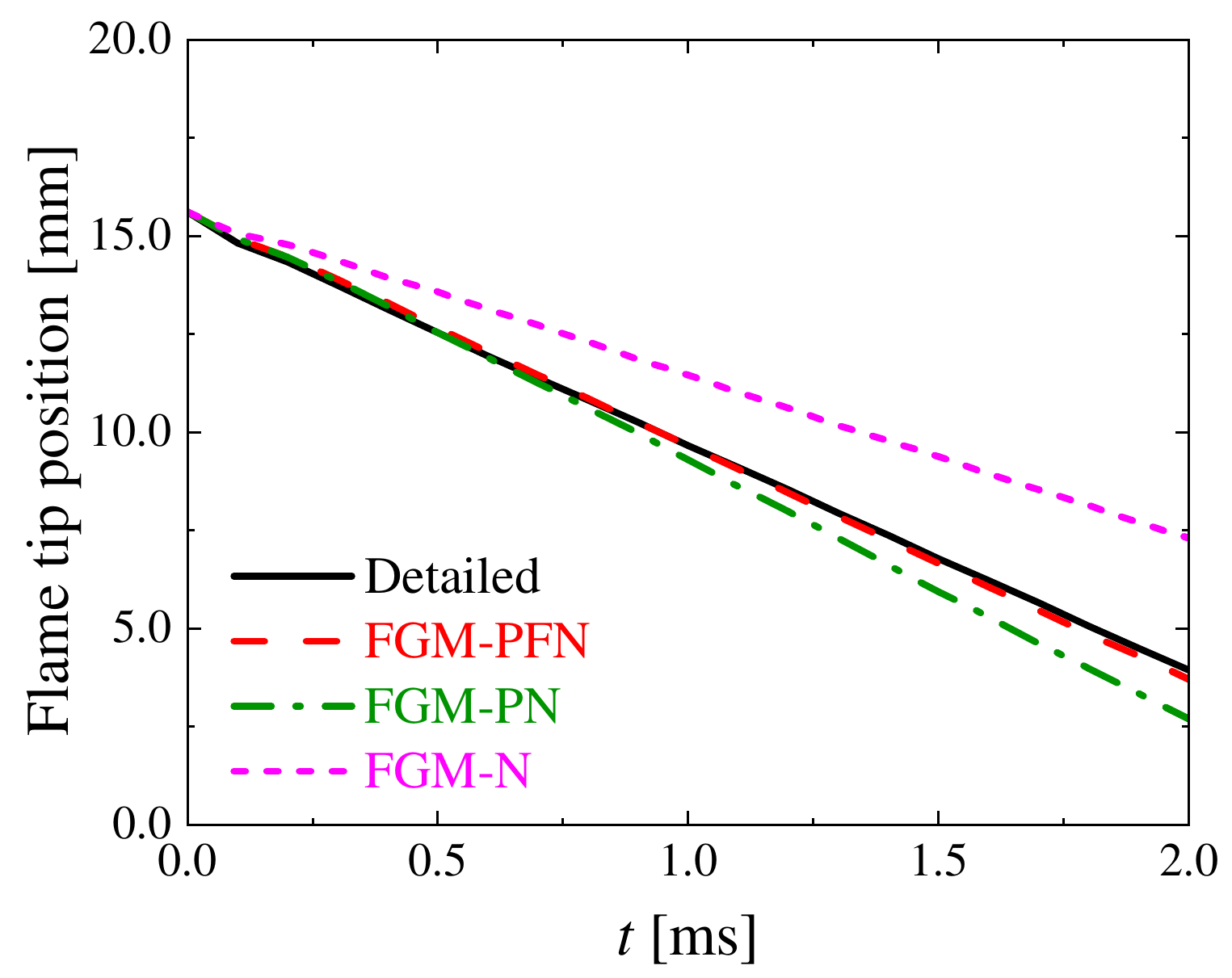}
					\subcaption{Flame tip position}
				\end{minipage} & \\
				\begin{minipage}[t]{0.7\hsize}
					\centering
					\includegraphics[width=1.0\linewidth]{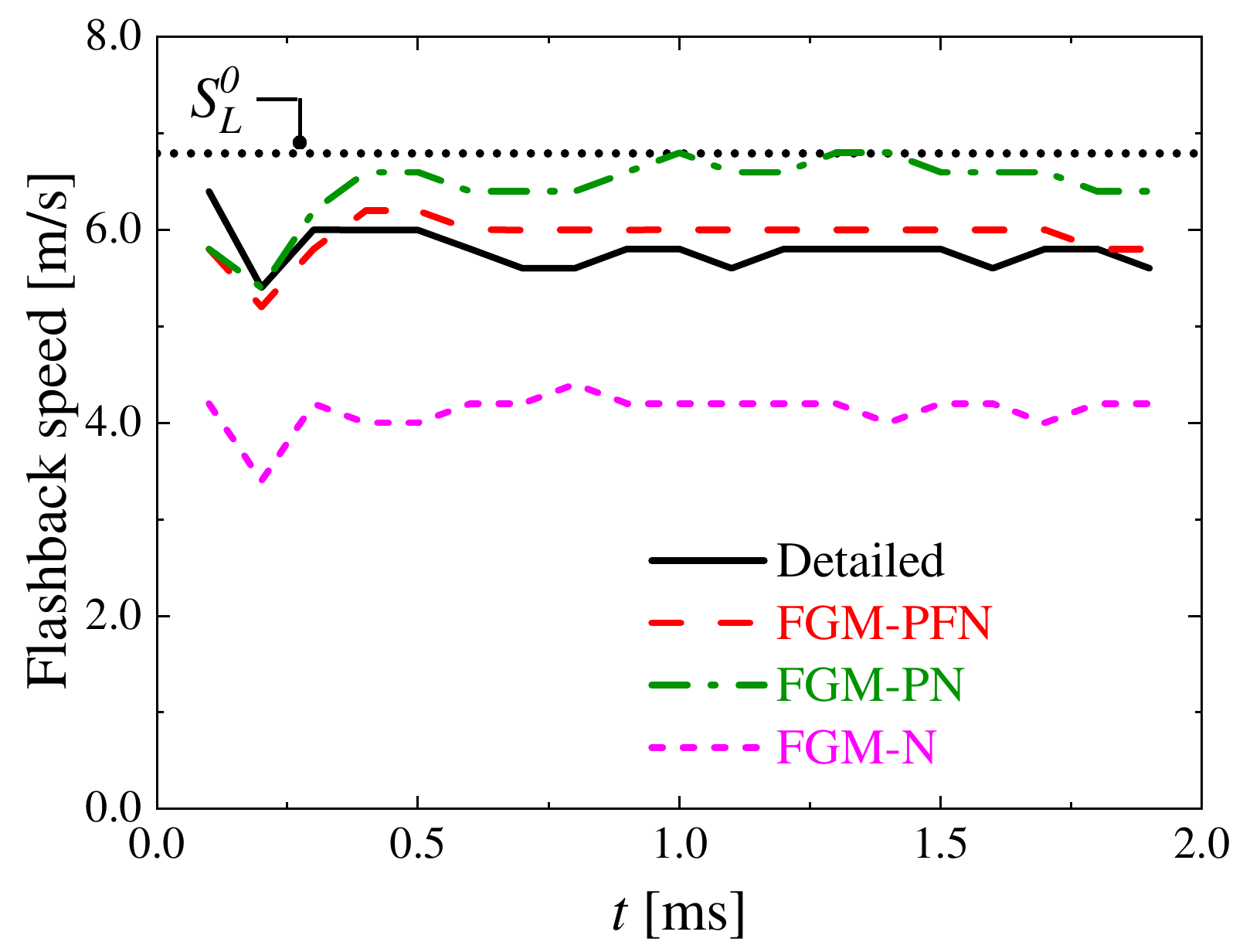}
					\subcaption{Flashback speed}
					\vspace{4 pt}
				\end{minipage}
			\end{tabular}\\
			\caption{Comparison of time variations of (a) flame tip positions and (b) flashback speed between detailed calculation, FGM-PFN, FGM-PN, and FGM-N methods. (b) A dotted black line indicates unstretched laminar burning velocity $S_{L}^0$.}		
			\label{fig:pos_speed}
		\end{figure}

\begin{figure}[htbp]
			\centering
			\begin{tabular}{cc}
				\centering
				\begin{minipage}[t]{0.7\hsize}
					\centering
					\includegraphics[width=1.0\linewidth]{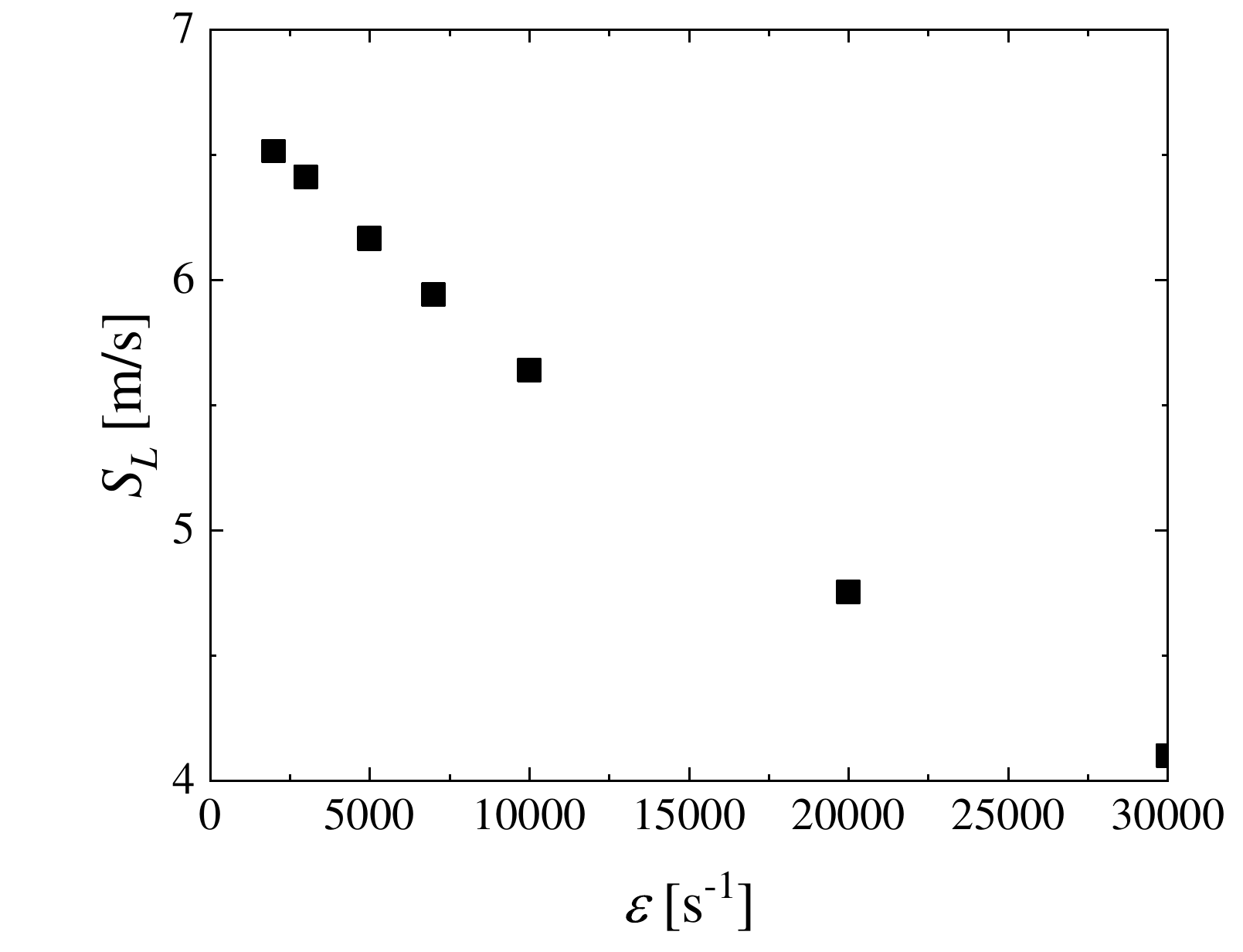}
					\subcaption{Effect of flame stretch rate}
					\label{fig:SL_FS}
				\end{minipage} & \\
				\begin{minipage}[t]{0.7\hsize}
					\centering
					\includegraphics[width=1.0\linewidth]{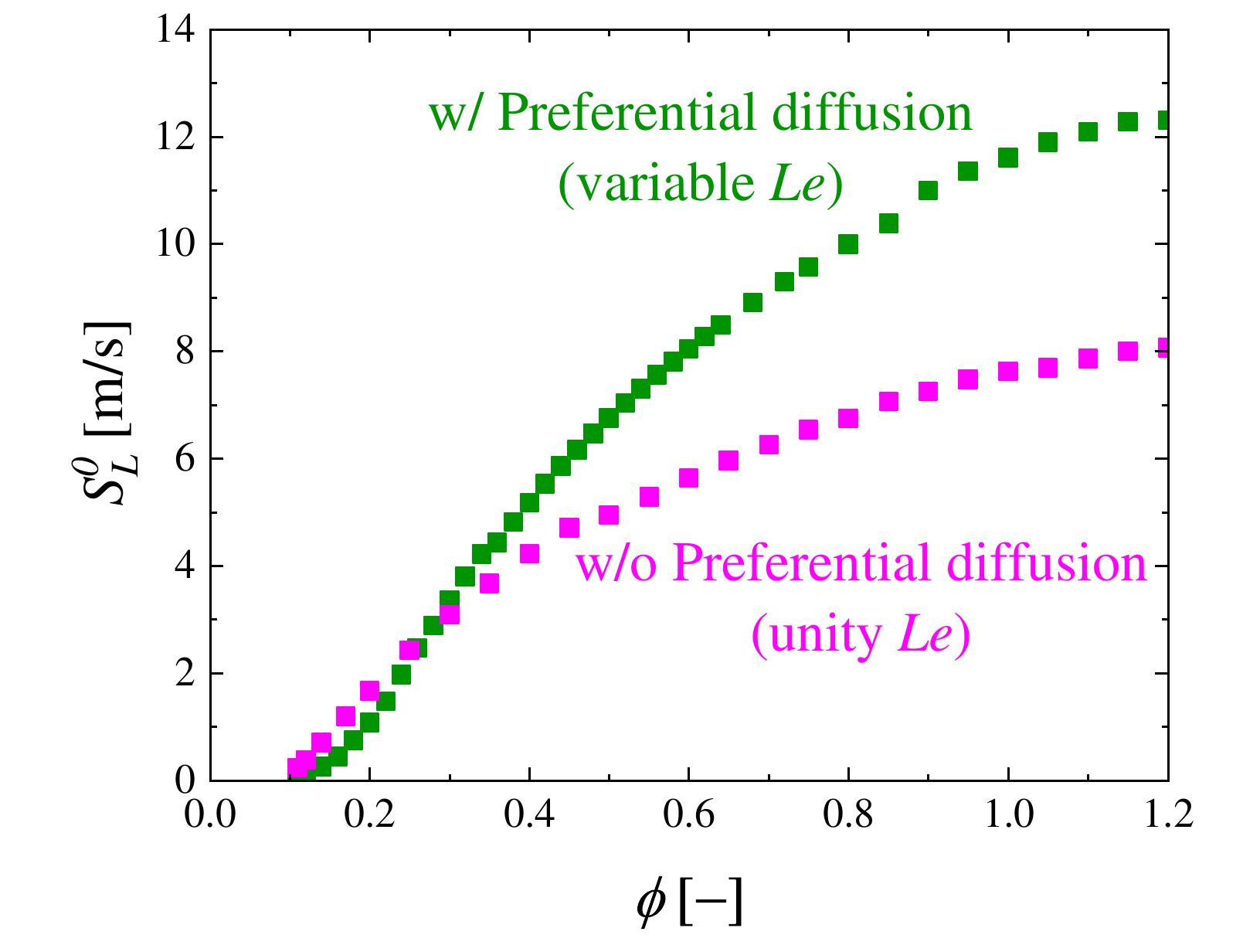}
					\subcaption{Effect of Lewis numbers}
					\label{fig:SL_PD}
					\vspace{3 pt}
				\end{minipage}
			\end{tabular}\\
			\caption{Effects of (a) flame stretch rate $\varepsilon$ and (b) Lewis numbers $Le$ on Laminar burning velocity $S_L$ under a pressure of \SI{1}{atm} and an unburnt temperature of \SI{750}{K} calculated by FlameMaster code. (a) $S_L$ versus $\varepsilon$ at an equivalence ratio $\phi$ of 0.5. (b) Unstretched laminar burning velocity $S_L^0$ versus $\phi$ obtained employing variable and constant $Le$.}		
			\label{fig:SL_PD_FS}
		\end{figure}

\begin{figure}[h!]
			\centering
			\includegraphics[width=0.7\linewidth]{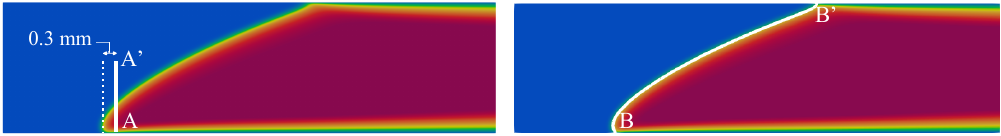}
			\vspace{2 pt}
			\centering
			\begin{tabular}{cc}
				\centering
				\begin{minipage}[t]{0.6\hsize}
					\centering
					\includegraphics[width=1.0\linewidth]{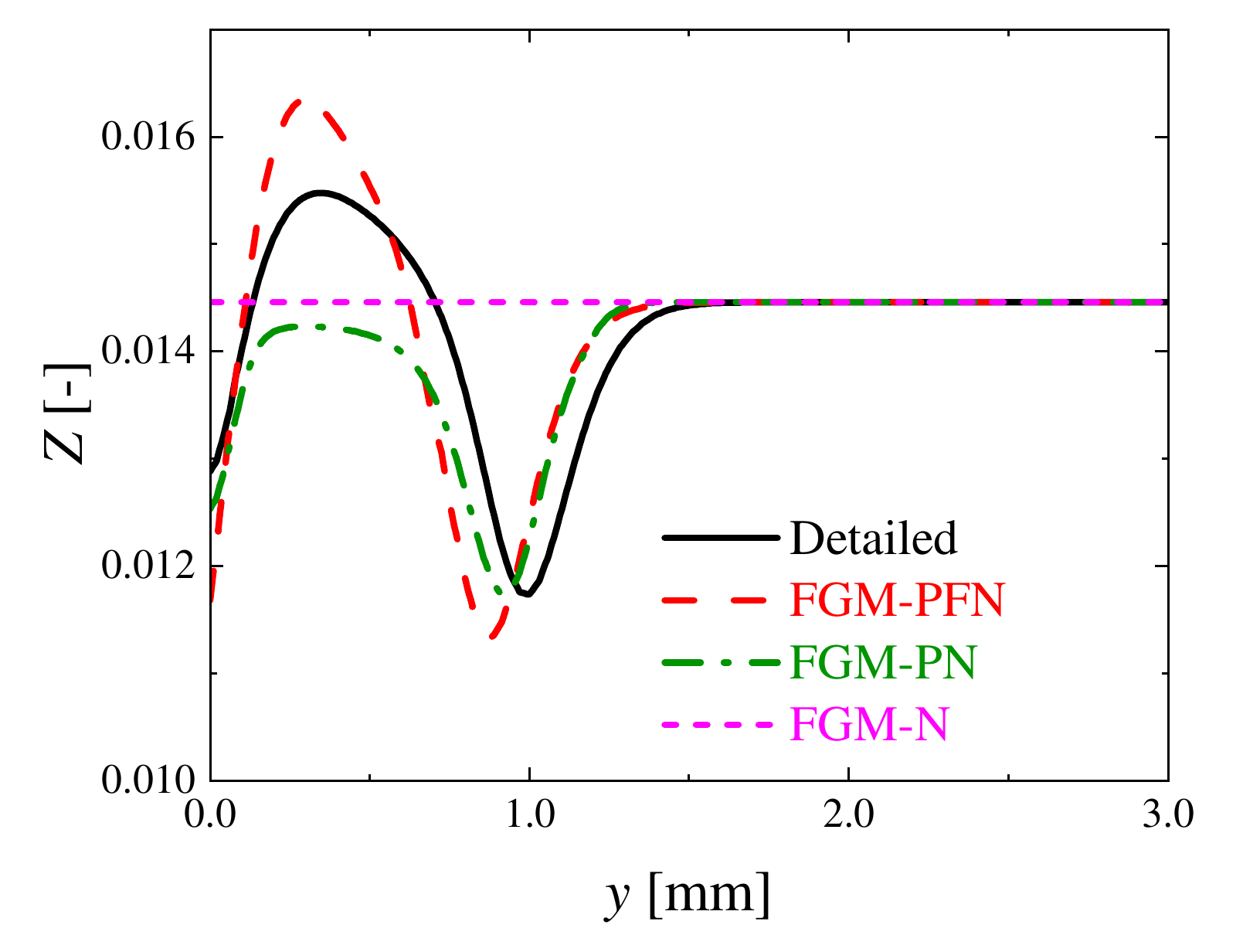}
					\subcaption{A-A'}
					\label{fig:Z_AA}
				\end{minipage} & \\
				\begin{minipage}[t]{0.6\hsize}
					\centering
					\includegraphics[width=1.0\linewidth]{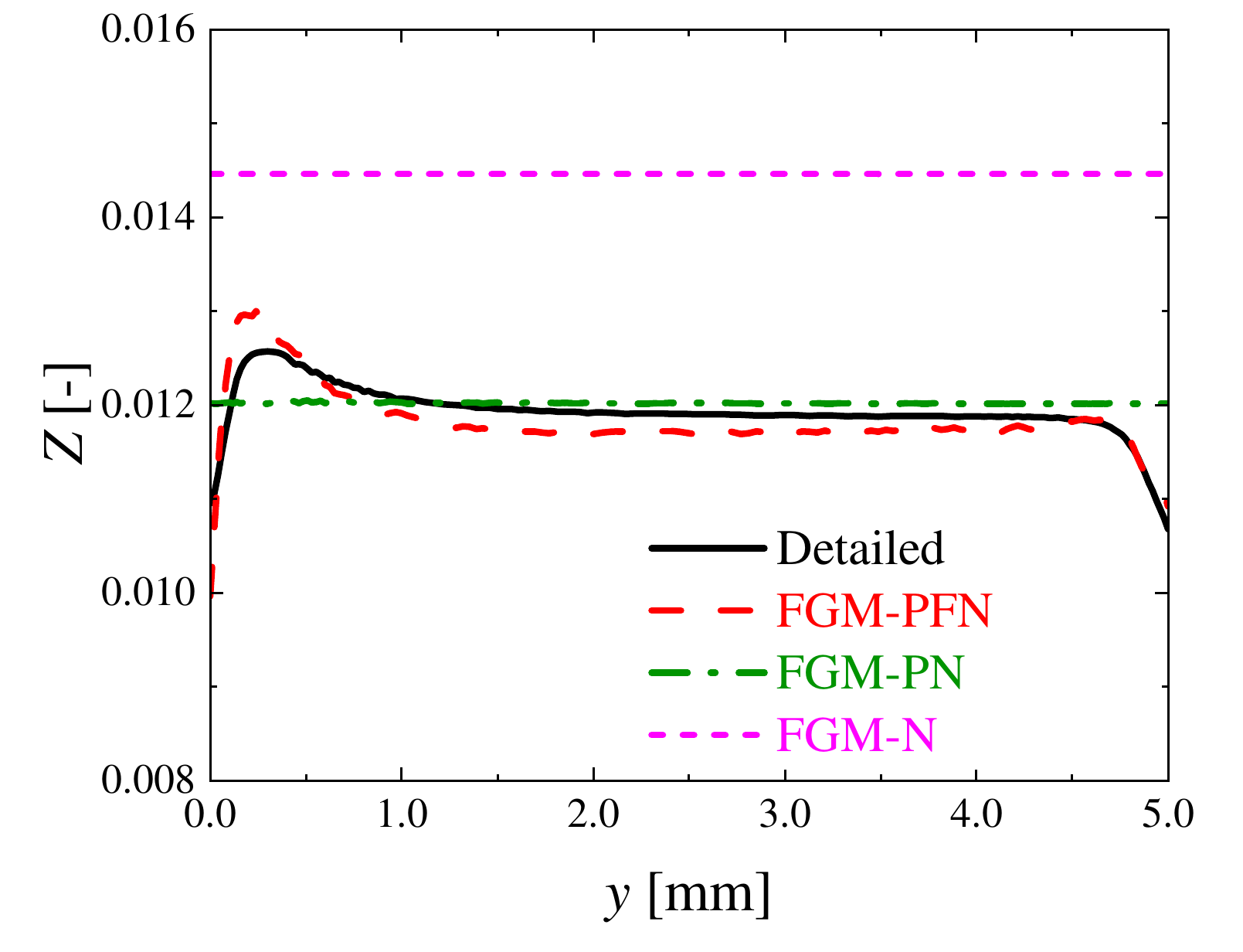}
					\subcaption{B-B'}
					\label{fig:Z_BB}
				\end{minipage}
			\end{tabular}\\
			\caption{Distributions of mixture fraction $Z$ across (a) a line A-A' and (b) a curve B-B’ (along the flame surface) obtained from detailed calculation, FGM-PFN, FGM-PN, and FGM-N methods at $t$ = \SI{2}{ms}.}	
			\label{fig:Z_AABB}
		\end{figure}

		Figure \ref{fig:T_field} shows the distributions of temperature $T$ obtained using the four methods.
		Figure \ref{fig:pos_speed} shows the time variations in the flame tip position and flashback speed.
		The flame tip position is the position of the flame front tip.
		In this section, the flame front is defined as the isosurface of the normalized progress variable $C_{norm}$ = 0.5.
		The flashback speed is defined as the propagation speed of the flame tip position.
		Figures \ref{fig:T_field} and \ref{fig:pos_speed} show that the FGM-PN method overestimates the flashback speed because it neglects the FS effect, whereas the FGM-N method underestimates it by neglecting the PD effect.
		The PD and FS effects in the one-dimensional \ce{H2}-air premixed flames are shown in Fig.~\ref{fig:SL_PD_FS}.
		The results in the figure are obtained from the one-dimensional numerical simulations of \ce{H2}-air premixed flames at various flame stretch rates using the FlameMaster code \cite{flamemaster}.
		The plotted results are used to generate the flamelet database.
		Figure \ref{fig:SL_FS} shows that the laminar burning velocity $S_L$ decreases with increasing flame stretch rate under the current conditions.
		This is consistent with the results of the FGM-PN method, which neglects the FS effect and overestimates the flashback speed, as shown in Fig.~\ref{fig:pos_speed}.
		Figure \ref{fig:SL_PD} shows that the unstretched laminar burning velocity $S_L^0$ decreases by neglecting the PD effect under an equivalence ratio $\phi$ = 0.5.
		This is consistent with the fact that the FGM-N method, which neglects the PD effect, underestimates the flashback speed, as shown in Fig.~\ref{fig:pos_speed}.
		However, the flashback speed predicted by the FGM-PFN method agrees well with that predicted by the detailed calculation as shown in Figs.~\ref{fig:T_field} and \ref{fig:pos_speed}.
		The flashback speed obtained from the detailed calculation is lower than $S_L^0$ mainly because the burning velocity decreases owing to the positive flame stretch rate rather than the heat loss effect.
		At the flame front, the thermal boundary layer does not develop for the heat loss to affect the flame speed at the flame tip.
		Figure \ref{fig:Z_AA} shows the distributions of the mixture fraction $Z$ (i.e., the local equivalence ratio) in the $y$-direction near the flame tip.
		For an unstretched \ce{H2}-air premixed flame, $Z$ decreases near the flame front because of the PD effect in the normal direction of the flame.
		If the flame front is convex/concave to the unburnt gas side, $Z$ increases/decreases because of the stronger convergence/divergence of \ce{H2} diffusion to the flame front compared with that of other species, such as \ce{O2}.
		As the FGM-N method neglects the PD and FS effects, it cannot reproduce the decreases in $Z$ near the flame front and the variation in $Z$ depending on the flame curvature (flame stretch) and provides a constant $Z$ distribution thoroughout the domain.
		A similar tendency was observed in a previous study \cite{kai2023flamelet}.
		As the FGM-PN method neglects the FS effect, it cannot reproduce the increase in $Z$ at the position where the flame front is convex to the unburnt gas side, whereas it can reproduce the decrease in $Z$ near the flame front.
		In the detailed calculation and FGM-PFN methods, $Z$ decreases near the flame front and increases at the flame convex part of the flame.
		Figure \ref{fig:Z_BB} shows the $Z$ distributions along the flame fronts ($C_{norm}=0.5$) at $t$ = 2 ms.
		The FGM-N method provides a constant and higher $Z$ profile.
		The FGM-PN method also provides a constant $Z$ profile, but it is close to the results of the detailed calculation, except in the vicinity of the wall. 
		The consideration of the PD effects in the FGM method results in the reproduction of the decrease in $Z$ near the flame front as shown in Fig.~\ref{fig:Z_AA}. 
		However, neglecting the FS effect causes the underestimation and overestimation of $Z$ at the flame convex part and near the wall. 
		The FGM-PFN method reproduces the decrease in $Z$ near the wall and the increase in $Z$ near the flame tip, although small discrepancies still exist compared with the profile of the detailed calculation. 
		These results demonstrate the effectiveness of the FGM-PFN method in the numerical simulation of a premixed \ce{H2} flame.

		\subsection{Three-dimensional turbulent boundary layer flashback} \label{subsec_3D}
\begin{figure}[p]
			\begin{tabular}{cc}
				\begin{minipage}[htbp]{0.5\hsize}
					\begin{center}
						\includegraphics[scale=0.25]{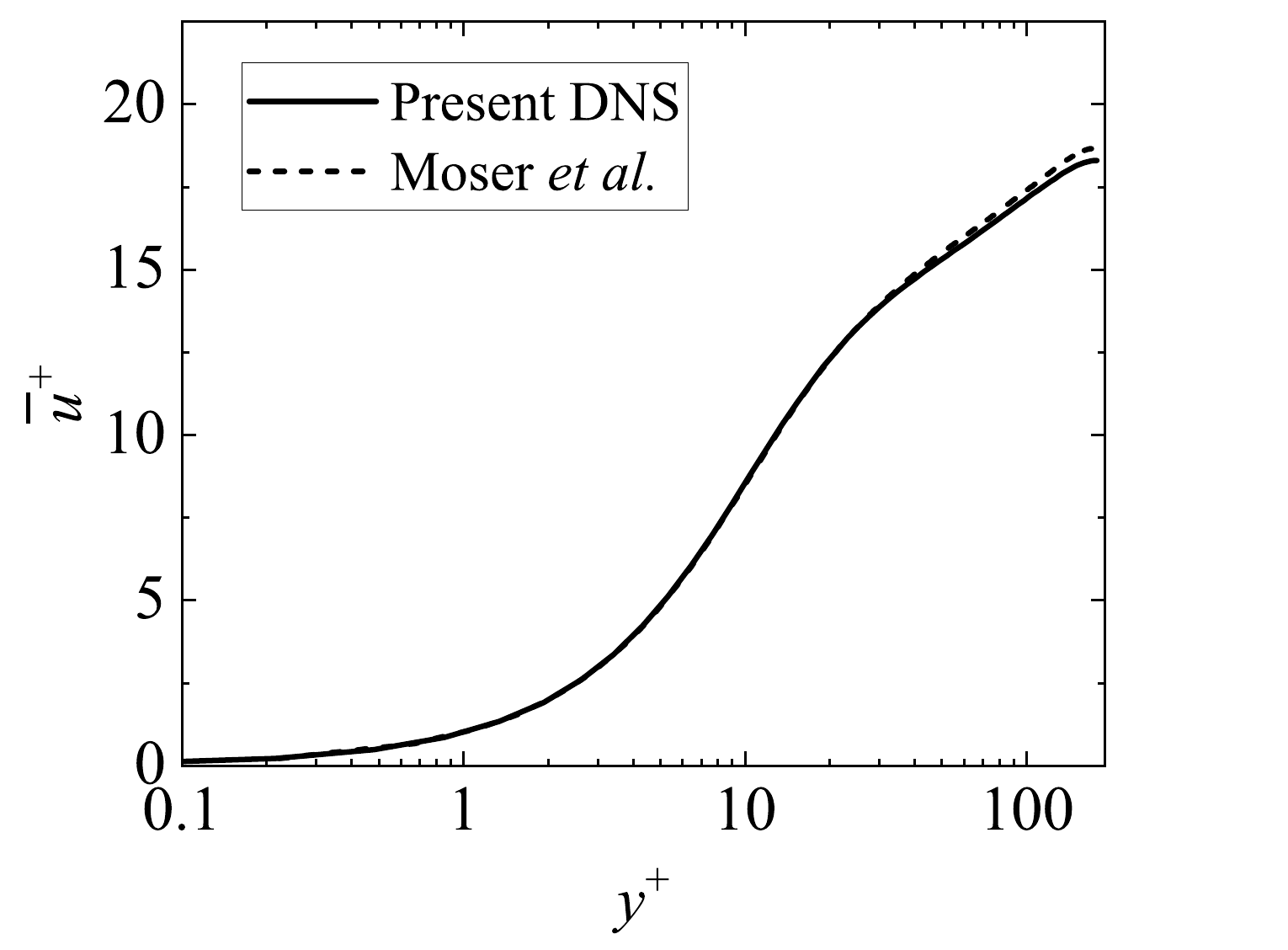}
					\end{center}
					\subcaption{$\bar{u}^+$}
					\label{up_mean}
				\end{minipage} &
				\begin{minipage}[htbp]{0.5\hsize}
					\begin{center}
						\includegraphics[scale=0.25]{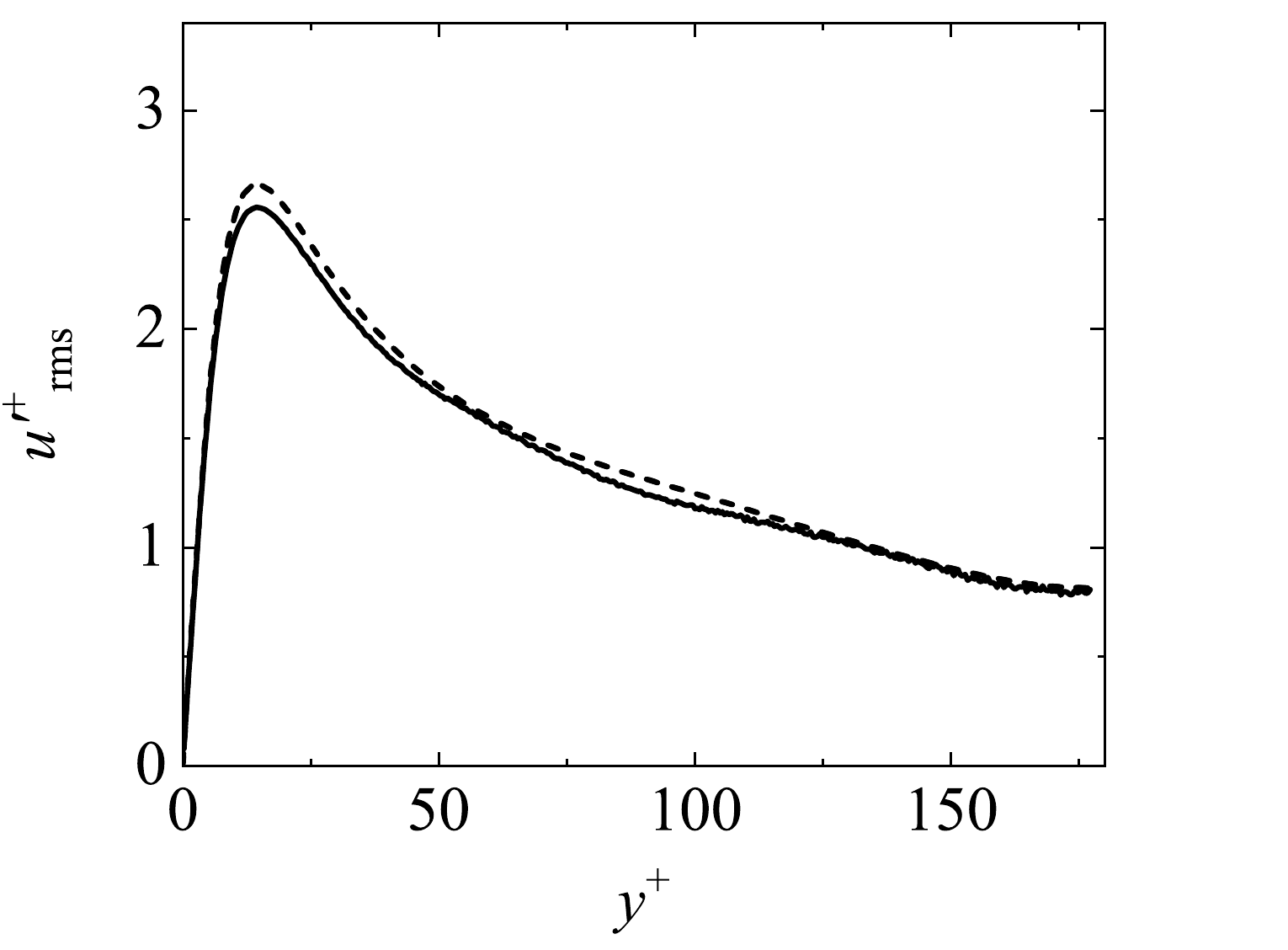}
					\end{center} 
					\subcaption{$u'$$^{+}_{rms}$}
					\label{up_rms}
				\end{minipage} \\
				
				\begin{minipage}[htbp]{0.5\hsize}
					\begin{center}
						\includegraphics[scale=0.25]{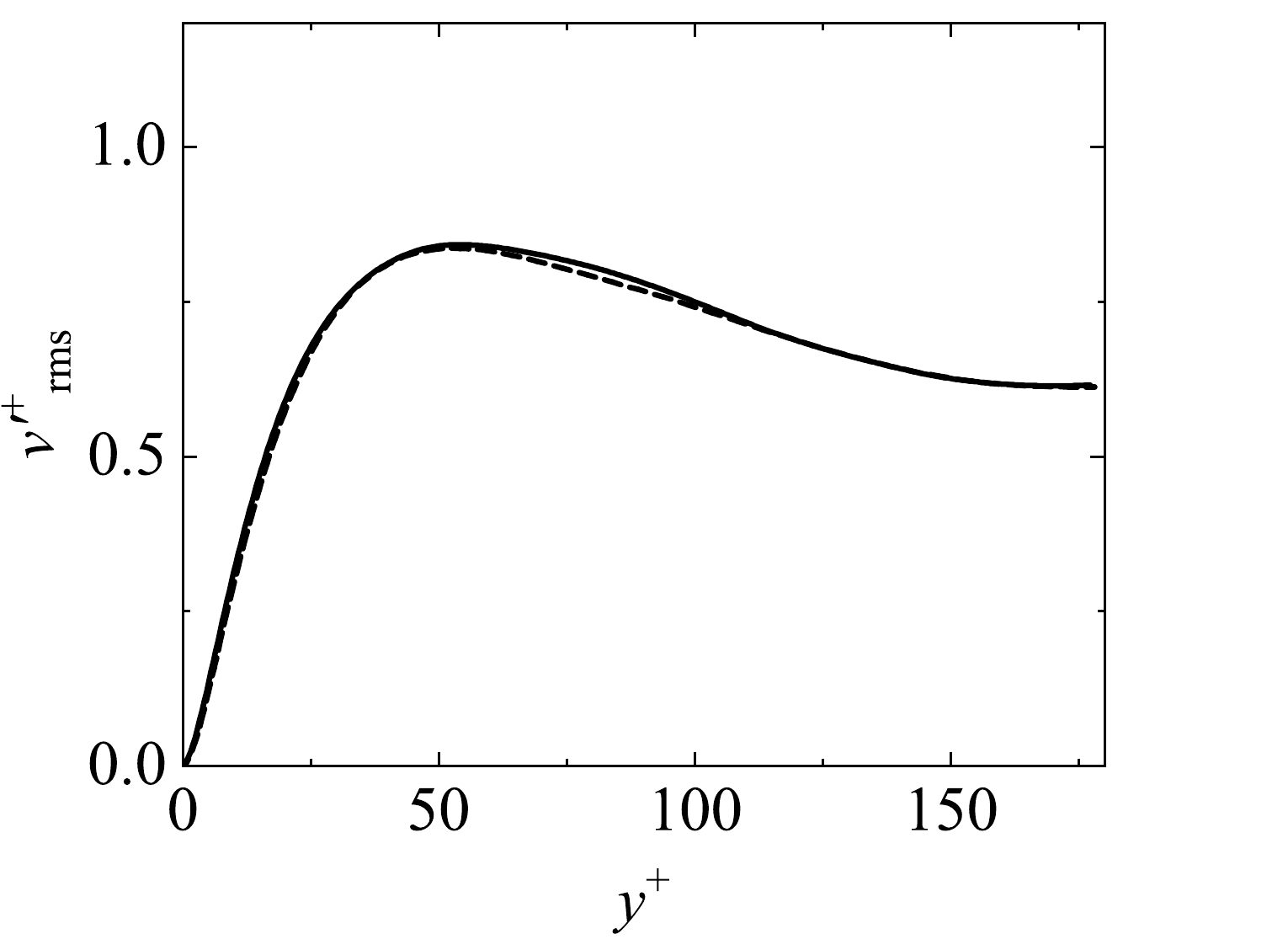}
					\end{center} 
					\subcaption{$v'$$^{+}_{rms}$}
					\label{vp_rms}
				\end{minipage} &
				\begin{minipage}[htbp]{0.5\hsize}
					\begin{center}
						\includegraphics[scale=0.25]{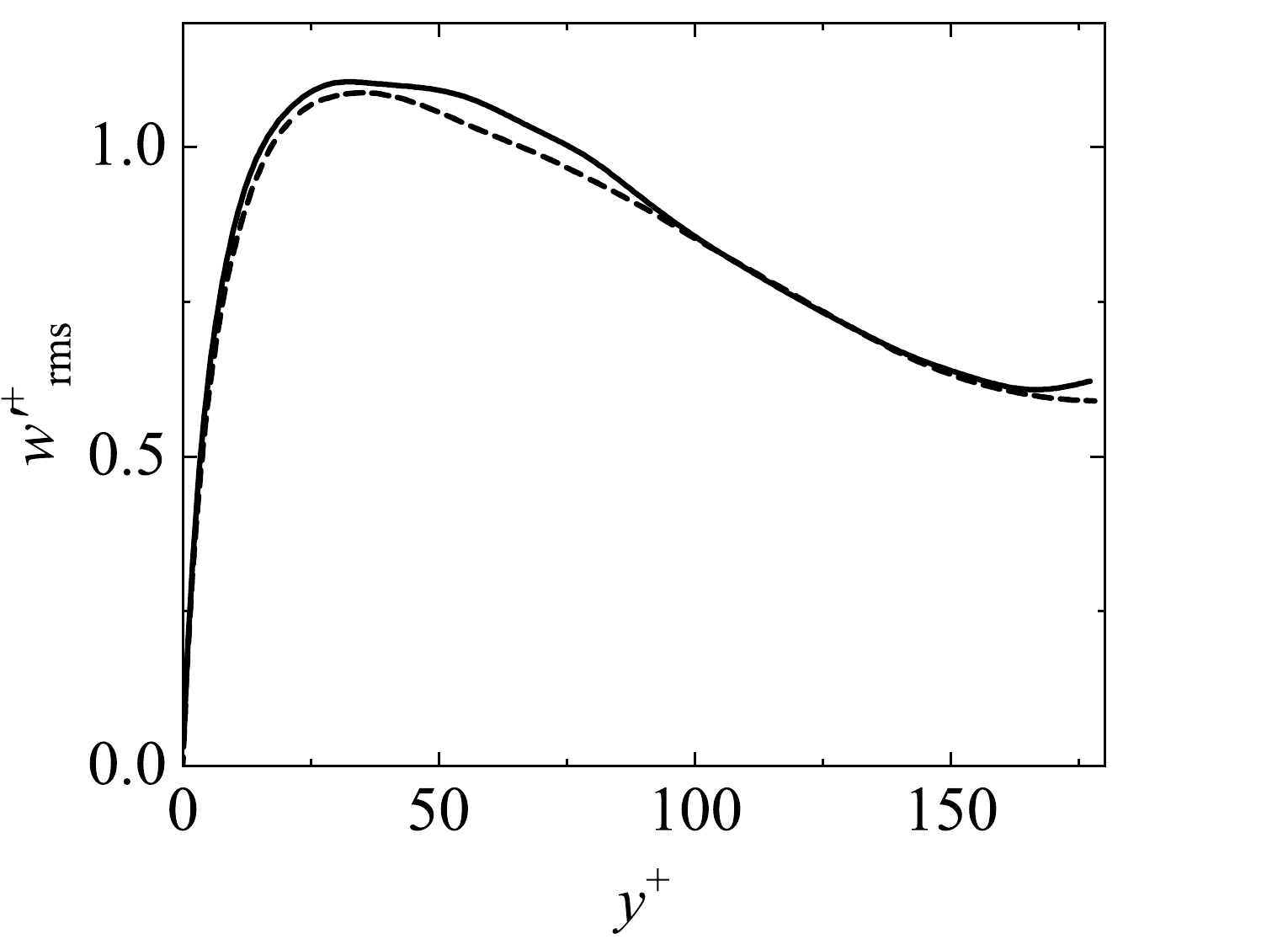}
					\end{center} 
					\subcaption{$w'$$^{+}_{rms}$}
					\label{wp_rms}
				\end{minipage} 
			\end{tabular}
			\caption{Profiles of (a) mean streamwise velocity, $\bar{u}^+$, and (b)-(d) RMS of velocity fluctuations, $u'$$^{+}_{rms}$,$v'$$^{+}_{rms}$,$w'$$^{+}_{rms}$, in wall normal direction obtained from 3D DNS of non-reacting flow in turbulence generation region.}
			\label{fig:stat}
		\end{figure}

		%
		Figure \ref{fig:stat} shows the profiles of several statistical properties: the mean streamwise velocity $\bar{u}^+$ and root mean square (RMS) of the velocity fluctuation in the $x$-, $y$-, and $z$-directions ($u'$$^{+}_{rms}$, $v'$$^{+}_{rms}$, and $w'$$^{+}_{rms}$).
		The results of a previous DNS proposed by Moser et al. \cite{moser1999direct} are also shown for comparison.
		The validity of wall-bounded turbulence in the present study is demonstrated because its statistical properties agree well with those of the previous DNS.
		
\begin{figure}[tbph]
			\centering
			\includegraphics[width=0.9\linewidth]{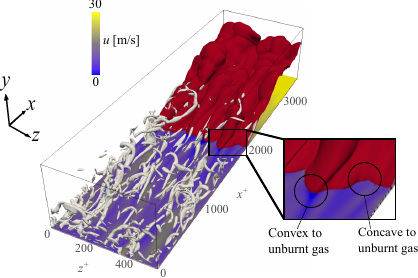}
			\caption{Instantaneous distribution of streamwise velocity $u$ on a $y^+$ = 2 plane, isosurface of temperature at 1000 K (red), and isosurface of the second invariant of velocity gradient tensor at $10^7$ $\rm{s^{-2}}$ (white) at $t$ = \SI{1.5}{ms} calculated by FGM-PFN method.}
			\label{fig:3Df_image}
		\end{figure}

\begin{figure}[tbph]
			\centering
			\includegraphics[width=0.7\linewidth]{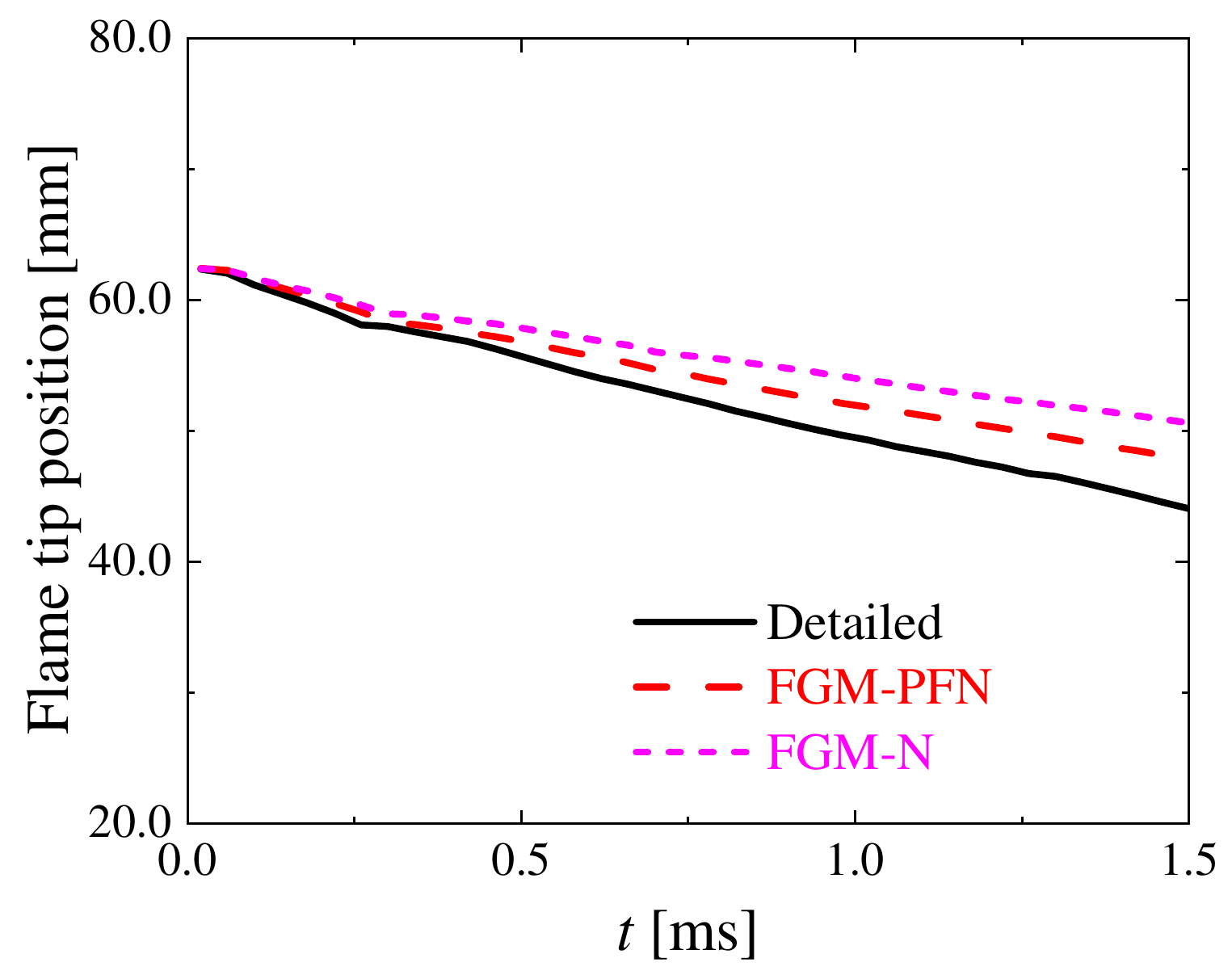}
			\caption{Time variation of flame tip positions obtained from detailed calculation, FGM-PFN, FGM-N methods.}
			\label{fig:3Df_pos}
		\end{figure}

		%
		Figure \ref{fig:3Df_image} shows an instantaneous image of the turbulent boundary layer flashback obtained using the FGM-PFN method.
		The streak structure in the wall turbulent flow induces flame wrinkling.
		This results in the formation of convex and concave parts, as shown in Fig.~\ref{fig:3Df_image}, which has also been observed in previous studies \cite{gruber2012direct,kitano2015effect}.
		Figure \ref{fig:3Df_pos} shows the time variations of the flame tip positions in the detailed calculation, and the FGM-PFN and FGM-N methods.
		The prediction accuracy of the flashback speed is improved using the FGM-PFN method compared with the FGM-N method, although a discrepancy from the detailed calculation still exists.
		In the following discussion, the reason for this discrepancy is examined by investigating the distributions of physical quantities in detail.

\begin{figure}[htbp]
			\centering
			\begin{tabular}{ccc}
				\centering
				\begin{minipage}[t]{1.0\hsize}
					\centering
					\includegraphics[width=1.0\linewidth]{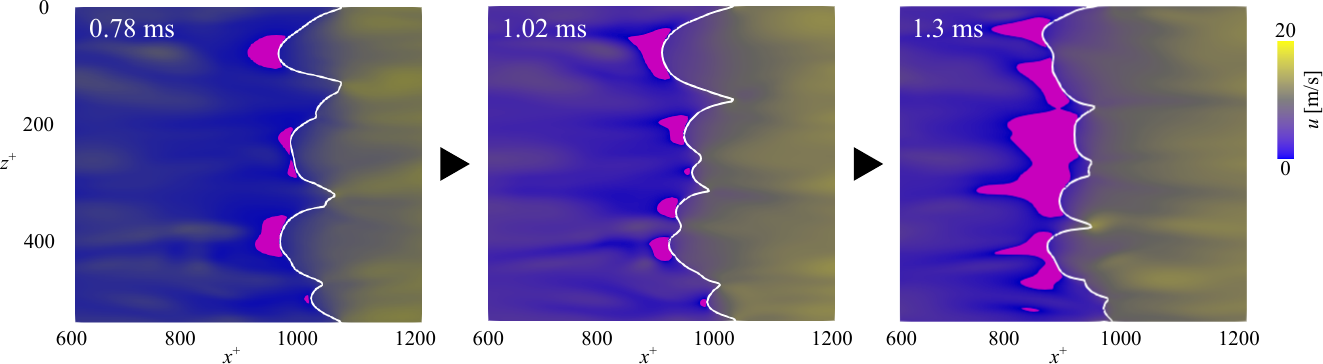}
					\subcaption{Detailed}
				\end{minipage} & \\
				\begin{minipage}[t]{1.0\hsize}
					\centering
					\includegraphics[width=1.0\linewidth]{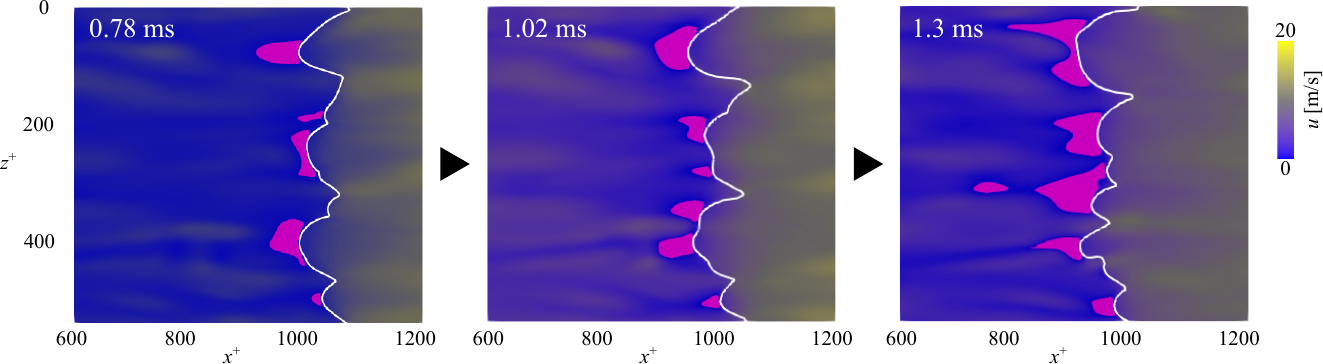}
					\subcaption{FGM-PFN}
				\end{minipage} & \\
				\begin{minipage}[t]{1.0\hsize}
					\centering
					\includegraphics[width=1.0\linewidth]{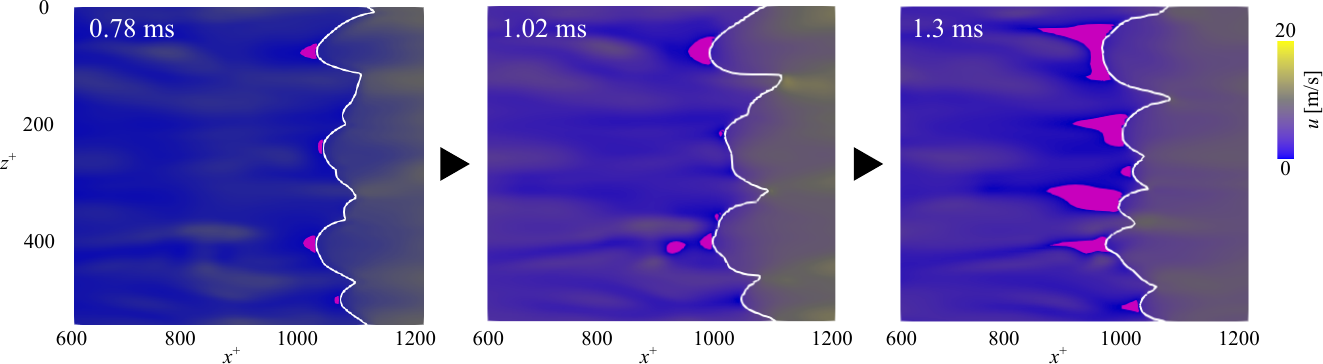}
					\subcaption{FGM-N}
				\end{minipage}
			\end{tabular}\\
			\caption{Distributions of streamwise velocity $u$ on $y^{+}$ = 2 plane at $t$ = \SI{0.78}{ms}, \SI{1.02}{ms}, and \SI{1.3}{ms} obtained from (a) detailed calculation, (b) FGM-PFN and (c) FGM-N methods. White solid lines show flame front defined as normalized progress variable $C_{norm}$ = 0.5 isosurface. Pink areas show backflow regions.}		
			\label{fig:u_on_y2}
		\end{figure}

		Figure \ref{fig:u_on_y2} shows sequential images of the distributions of streamwise velocity $u$ on the $y^+$ = 2 plane.
		The backflow regions, where the streamwise velocity is negative, exist immediately upstream of the part where the burnt area is convex to the unburnt area.
		The formation of backflow regions has also been reported in previous experiments \cite{eichler2012premixed} and DNSs \cite{gruber2012direct}.
		As shown in Fig.~\ref{fig:u_on_y2}, the backflow region predicted by the FGM-PFN method is similar to that predicted by the detailed calculation and larger than that predicted by FGM-N method.
		Therefore, the PD and FS effects are important, even in turbulent flames.
		According to Gruber et al. \cite{gruber2012direct}, flame wrinkling induced by the Darrieus--Landau instability plays an important role in the formation of the backflow region.
		PD and FS may also enhance flame wrinkling via thermodiffusive instability.
		However, in the present study, the flame shapes of the FGM-PFN and FGM-N methods are similar, which indicates that the effects of PD and FS on the flame shape are smaller than the effect of turbulence at the beginning of the flashback.
		%
		%
		%
%
		%
				
\begin{figure}[h!]
			\centering
			\begin{tabular}{ccc}
				\centering
				\begin{minipage}[t]{0.5\hsize}
					\centering
					\includegraphics[width=1.0\linewidth]{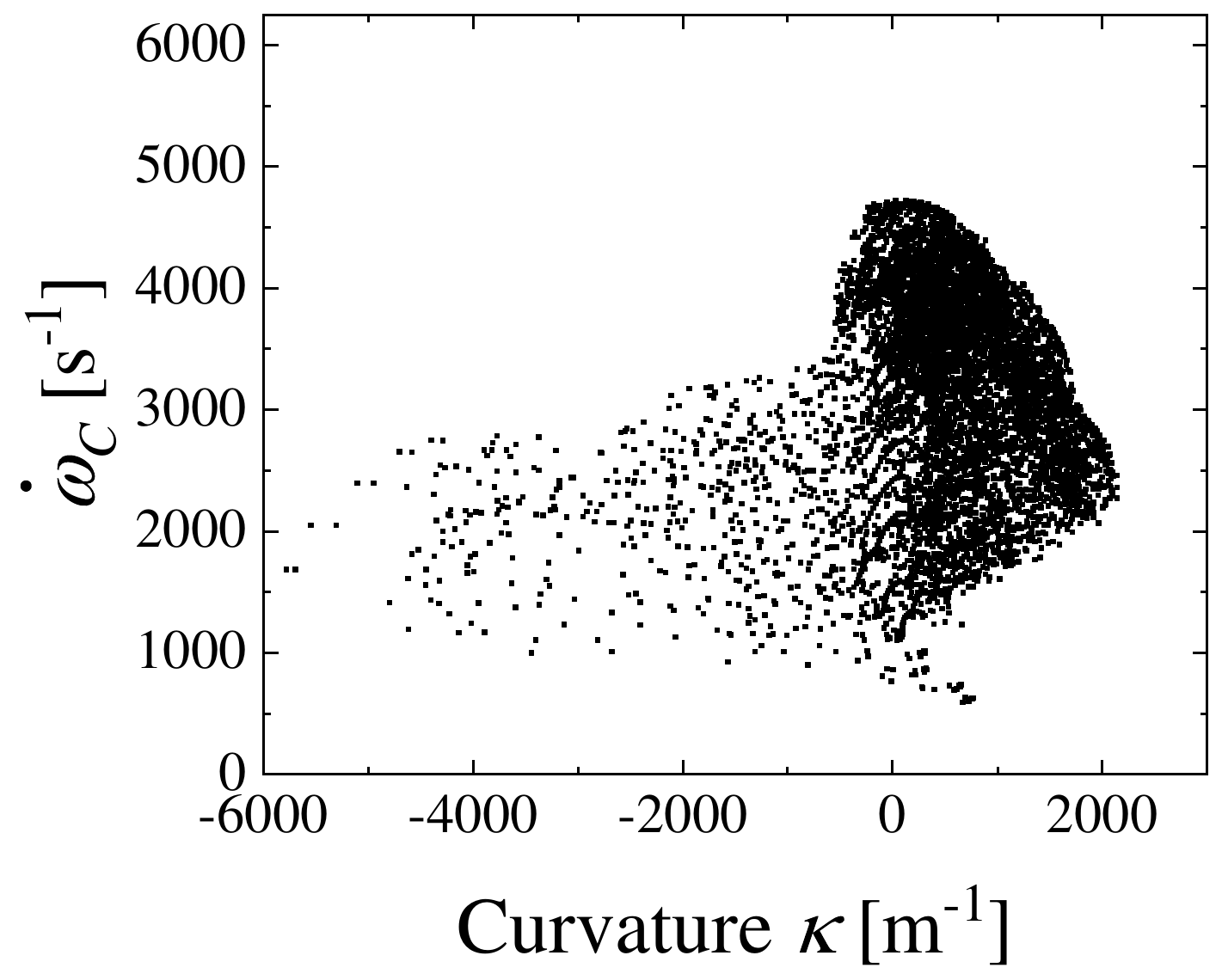}
					\subcaption{Detailed}
					\label{fig:wC_k_detail}
				\end{minipage} & \\
				\begin{minipage}[t]{0.5\hsize}
					\centering
					\includegraphics[width=1.0\linewidth]{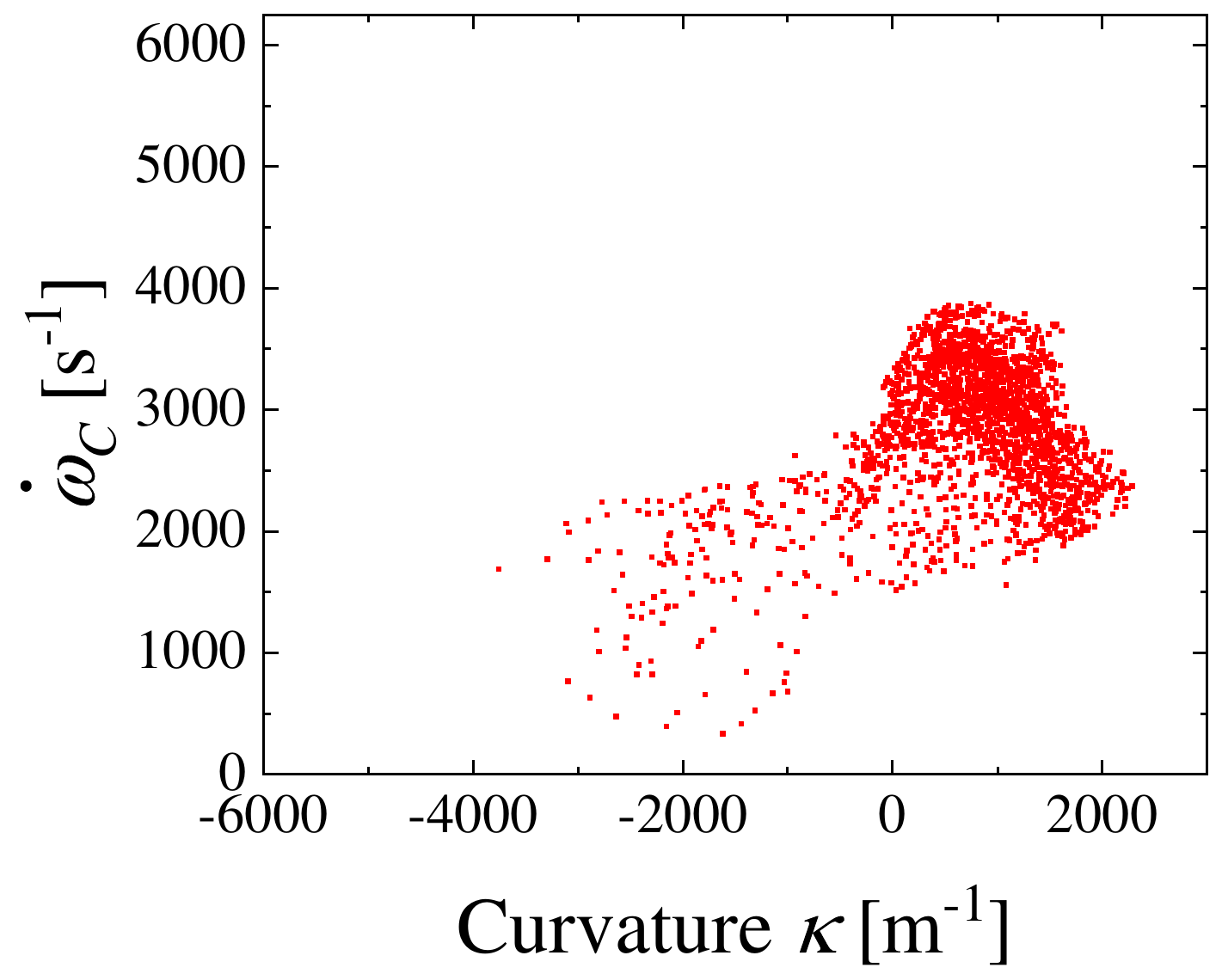}
					\subcaption{FGM-PFN}
					\label{fig:wC_k_PFN}
				\end{minipage} & \\
				\begin{minipage}[t]{0.5\hsize}
					\centering
					\includegraphics[width=1.0\linewidth]{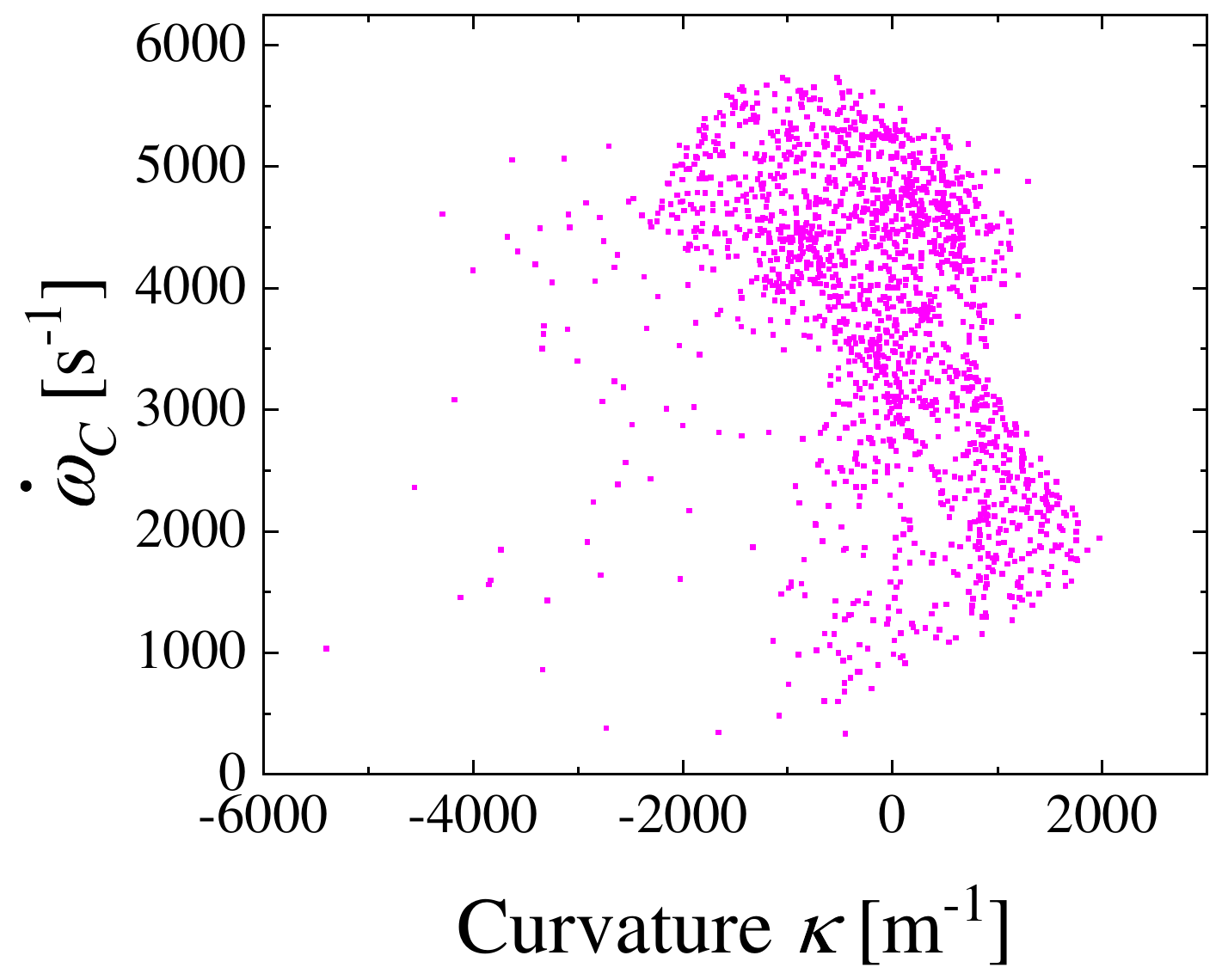}
					\subcaption{FGM-N}
					\label{fig:wC_k_N}
				\end{minipage}
			\end{tabular}\\
			\caption{Reaction rate of progress variable $C$, $\dot{\omega}_{C}$, plotted versus curvature $\kappa$ near the flame front on $y^{+}$ = 2 plane at $t$ = \SI{0.78}{ms} obtained from (a) detailed calculation, (b) FGM-PFN, and (c) FGM-N methods.}
			\label{fig:wC_vs_k}
		\end{figure}

\begin{figure}[h!]
\setkeys{Gin}{width=\linewidth}
\captionsetup{skip=0.5ex}
\begin{minipage}{.3\linewidth}
	\includegraphics{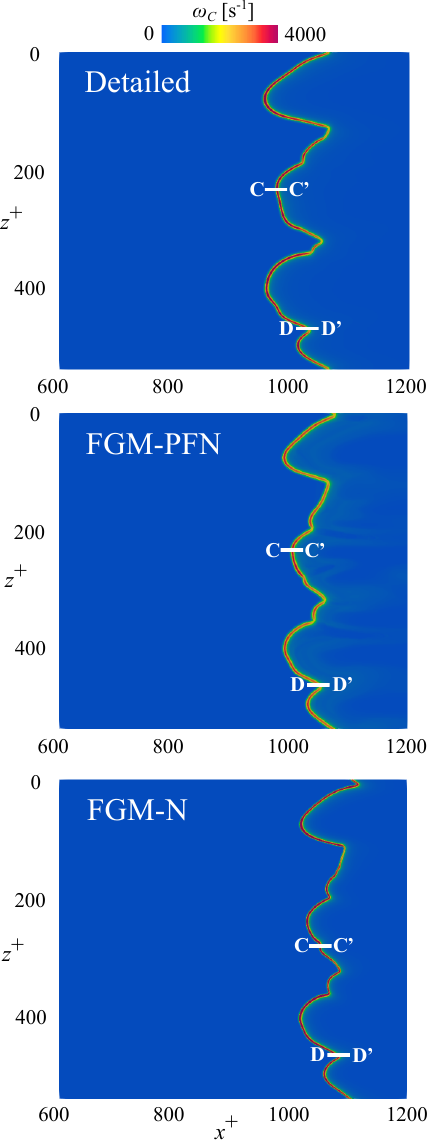}
	\subcaption*{Compared flame front}
\end{minipage}\hfill
\begin{minipage}{.65\linewidth}
	\includegraphics{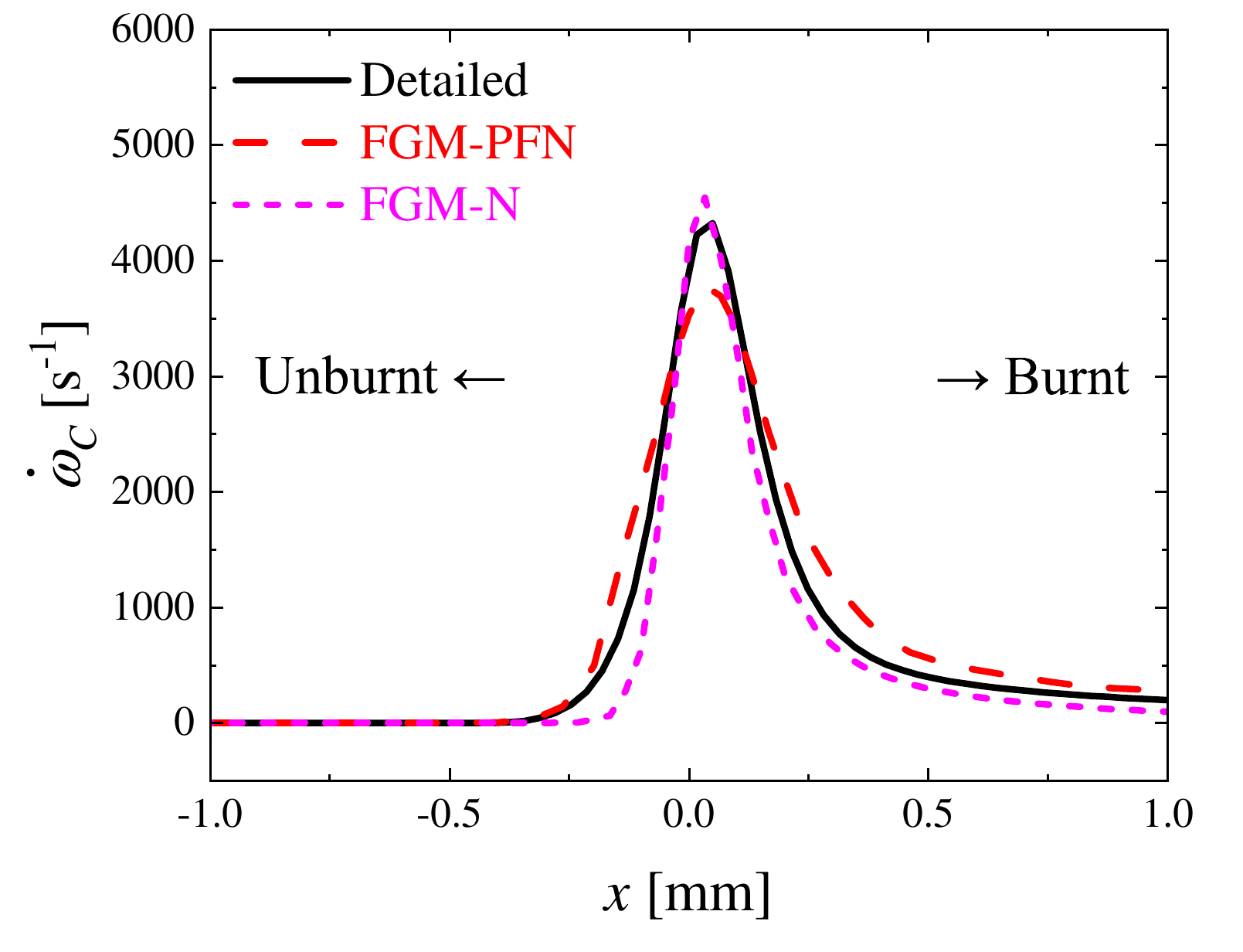}
	\subcaption{C-C' (convex to the unburnt side)}
	\label{fig:wC_convex}
	
	\bigskip
	\includegraphics{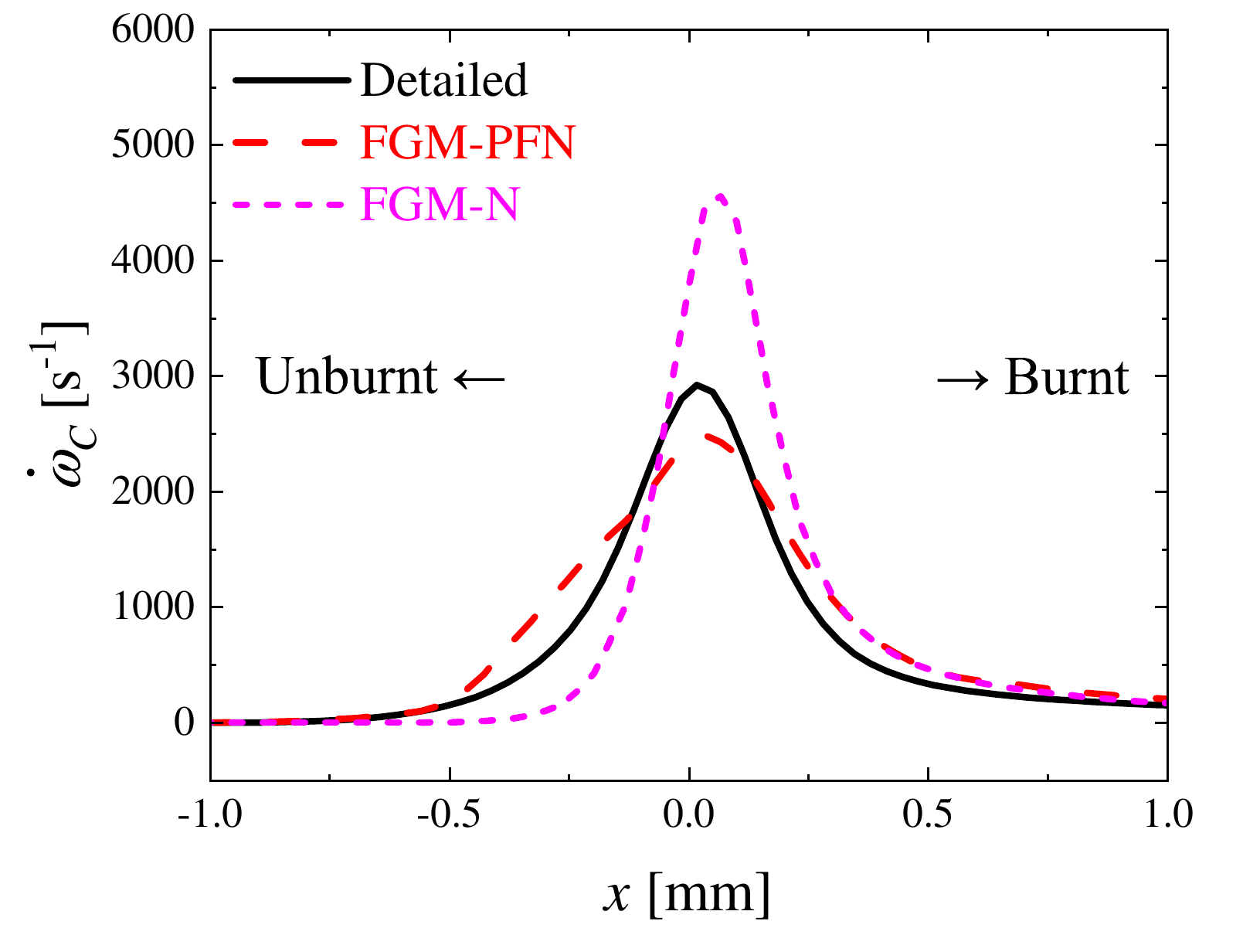}
	\subcaption{D-D' (concave to the unburnt side)}
	\label{fig:wC_concave}
\end{minipage}\hfill

			\caption{Distributions of reaction rate of progress variable $C$, $\dot{\omega}_{C}$, across (a) C-C' (convex to the unburnt side) and (b) D-D' (concave to the unburnt side) at $t$ = \SI{0.78}{ms} obtained from detailed calculation, FGM-PFN, and FGM-N methods. Flame front, in which normalized $C$, $C_{norm}$ = 0.5, is at $x$ = \SI{0}{mm}.}
			\label{fig:wC_vs_x}
		\end{figure}

		%
		Figure \ref{fig:wC_vs_k} shows $\dot{\omega}_C$ (= $\dot{\omega}_{\ce{H2O}}$) plotted versus curvature $\kappa$ on the $y^+$ = 2 plane near the flame front at $t$ = \SI{0.78}{ms}.
		$\kappa$ is written as:
		\begin{equation} \label{eq:kappa}
			\kappa = \frac{1}{2} \nabla \cdot \bm{n},
		\end{equation}
		\begin{equation} \label{eq:n}
			\bm{n} = - \frac{\nabla C}{|\nabla C|}.
		\end{equation}
		where $\bm{n}$ denotes the unit vector normal to the flame front.
		The results of the detailed calculation and the FGM-PFN method show higher $\dot{\omega}_C$ at the convex flame front ($\kappa >$ 0) and lower $\dot{\omega}_C$ at the concave flame front ($\kappa <$ 0).
		However, the $\dot{\omega}_C$ reproduced by the FGM-N method does not vary much between the convex and concave areas.
		To investigate the effects of PD and FS on $\dot{\omega}_C$ in more detail, Fig.~\ref{fig:wC_vs_x} shows the comparisons of the spatial distributions of $\dot{\omega}_C$ across the convex and concave flame fronts on the $y^+$ = 2 plane.
		The locations of the compared flame fronts (convex flame fronts C-C' and concave flame fronts D-D') are also shown in the figure.
		These compared flame fronts of each numerical method are determined so that the $\kappa$ of each flame has a similar value.
		The $\kappa$ value at $C_{norm}$ = 0.5 on the flame front C-C' is \SI{690}{s^{-1}} for the detailed calculation, \SI{670}{s^{-1}} for the FGM-PFN method, and \SI{700}{s^{-1}} for the FGM-N method.
		The $\kappa$ value at $C_{norm}$ = 0.5 on the flame front D-D' is \SI{-1220}{ s^{-1}} for the detailed calculation, \SI{-1200}{s^{-1}} for the FGM-PFN method, and \SI{-1210}{s^{-1}} for the FGM-N method.
		The figure shows that the FGM-N method produces similar $\dot{\omega}_C$ distributions at the convex and concave flame fronts.
		On the other hand, the low $\dot{\omega}_C$ peak at the concave flame front is successfully reproduced in the FGM-PFN method and in the detailed calculation.
		Regardless of the convex or concave flame fronts, the peak values of $\dot{\omega}_C$ are larger in the FGM-N method than in the detailed calculation.
		This is because the local equivalence ratio of the FGM-N method is closer to the stoichiometry than that of the detailed calculation, which is similar to the $Z$ profile shown in Fig.~\ref{fig:Z_BB}.
		However, for both flame fronts, the $\dot{\omega}_C$ of the detailed calculation is larger than that of the FGM-N method in the region closer to the unburnt side.
		This may lead to the underestimation of the flashback speed of the FGM-N method in spite of the higher peak values of $\dot{\omega}_C$.
		On the other hand, the peak values of $\dot{\omega}_C$ are slightly smaller in the FGM-PFN method than in the detailed calculation.
		This may lead to the underestimation of the flashback speed in the FGM-PFN method.
		A comparison of the results of the detailed calculation in Figs.~\ref{fig:wC_convex} and \ref{fig:wC_concave} further shows that the convex and concave shapes of the flame affect where the reaction occurs in the flame zone.
		In other words, $\dot{\omega}_C$ is relatively high near the burnt side across the convex flame front and high near the unburnt side across the concave flame front.
		$Z$ increases on the burnt side of the convex part and decreases on the burnt side of the concave part in premixed \ce{H2} flames \cite{regele2013two}.
		Accordingly, $Z$ is higher on the burnt side across the convex flame front and on the unburnt side across the concave flame front, and $\dot{\omega}_C$ becomes higher where $Z$ is large (the local equivalence ratio is large), which is consistent with the results of Fig.~\ref{fig:wC_vs_x}.
		These results indicate that PD and FS cause the different trends in the $\dot{\omega}_C$ distributions of the convex and concave flames. 

		%
		%

\begin{figure}[htbp]
			\centering
			\begin{tabular}{cccc}
				\centering
				\begin{minipage}[t]{0.5\hsize}
					\centering
					\includegraphics[width=1.0\linewidth]{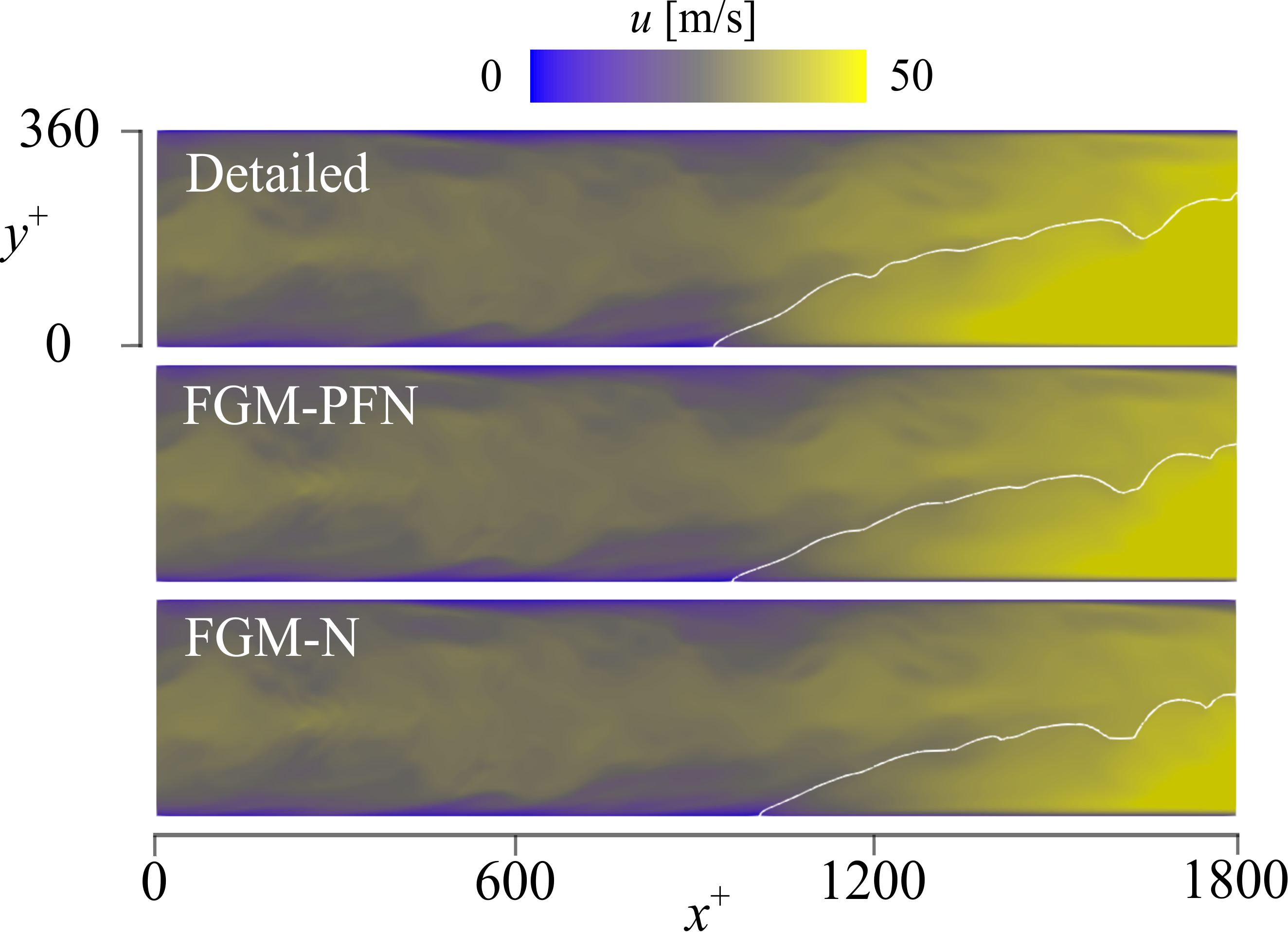}
					\subcaption{Streamwise velocity $u$}
				\end{minipage} & 
				\begin{minipage}[t]{0.5\hsize}
					\centering
					\includegraphics[width=1.0\linewidth]{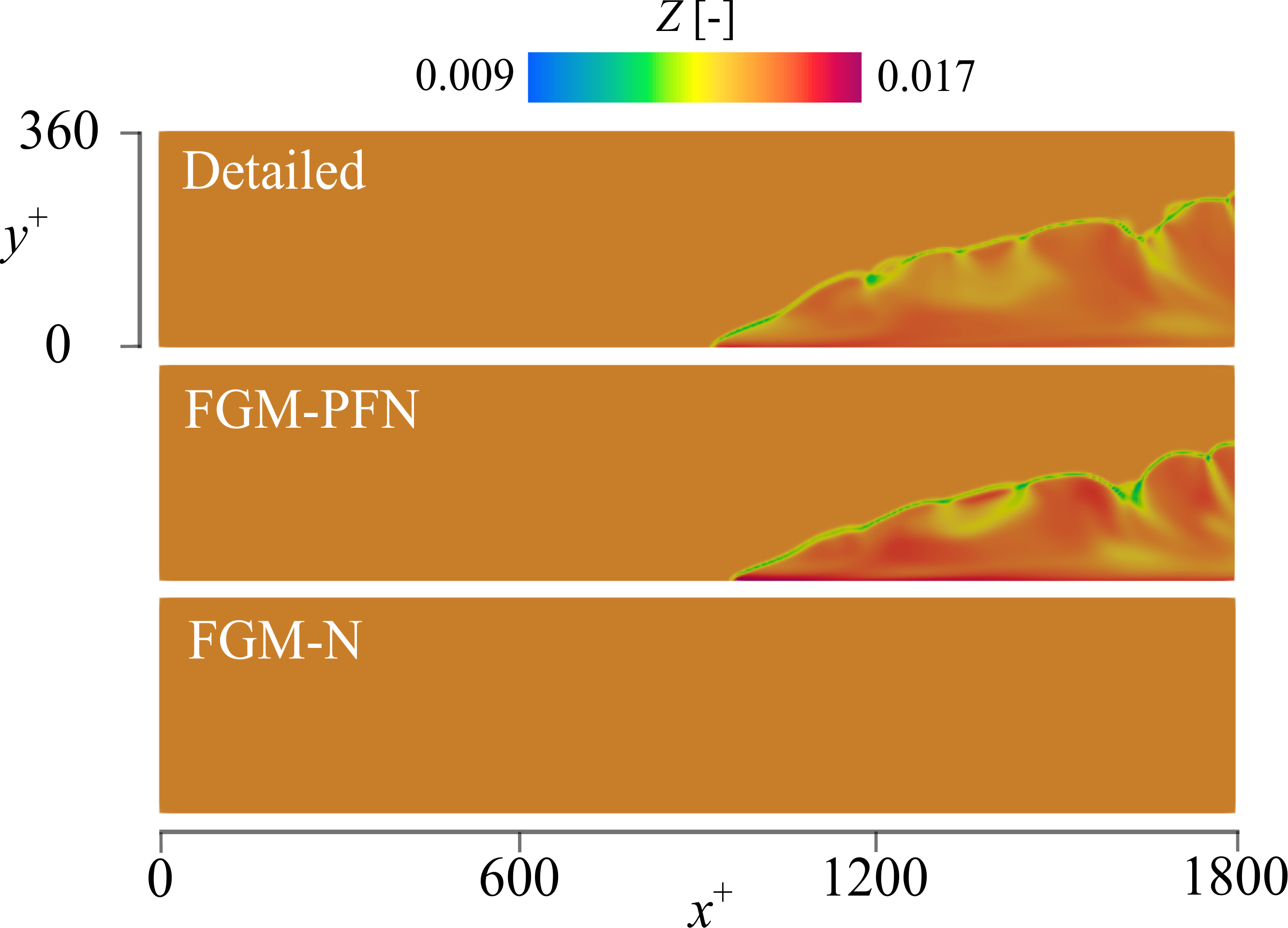}
					\subcaption{Mixture fraction $Z$}
				\end{minipage} & \\
				\begin{minipage}[t]{0.5\hsize}
					\centering
					\includegraphics[width=1.0\linewidth]{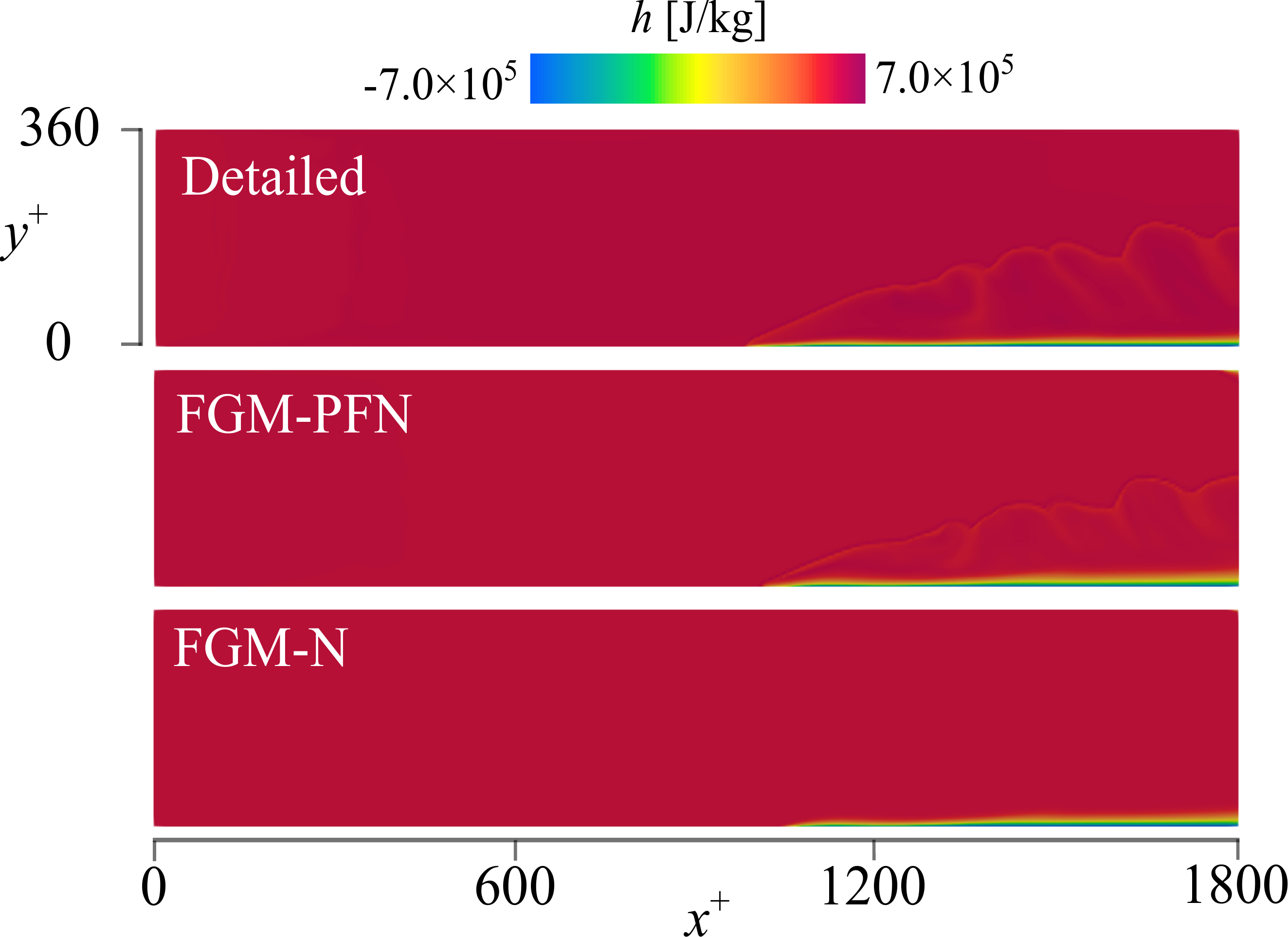}
					\subcaption{Specific enthalpy $h$}
				\end{minipage} & 
				\begin{minipage}[t]{0.5\hsize}
					\centering
					\includegraphics[width=1.0\linewidth]{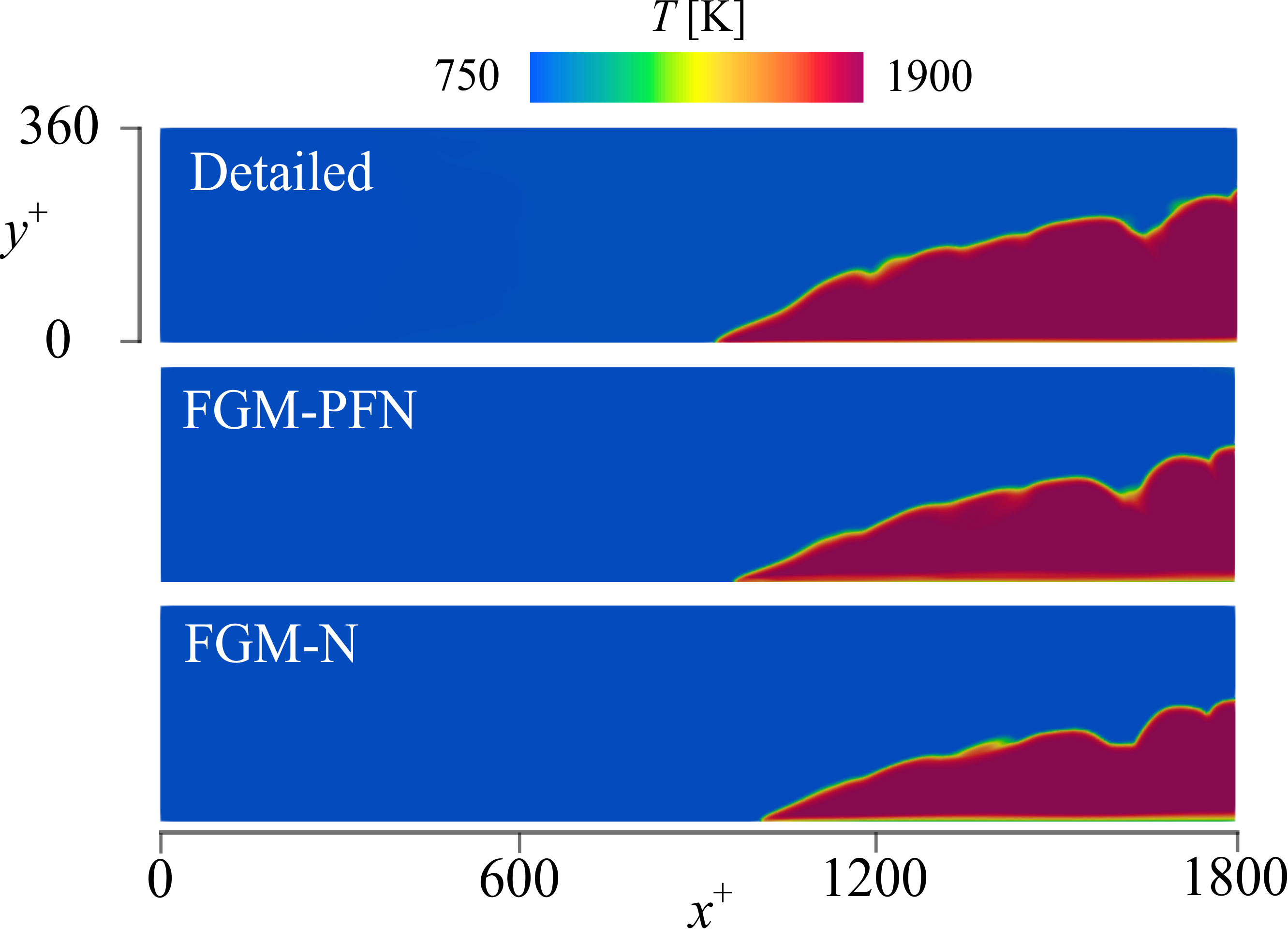}
					\subcaption{Temperature $T$}
				\end{minipage} \\
			\end{tabular}\\
			\caption{Instantaneous distributions of (a) streamwise velocity $u$, (b) mixture fraction $Z$, (c) specific enthalpy $h$, and (d) temperature $T$ at $t$ = \SI{1.3}{ms} on $z^+$ = 270 plane (center of the channel). White solid lines in (a) represent flame front.}
			\label{fig:z_direc}
		\end{figure}

		Figure \ref{fig:z_direc} shows the instantaneous distributions of the streamwise velocity $u$, mixture fraction $Z$, specific enthalpy $h$, and temperature $T$ in the $z^+$ = 270 plane.
		As shown in the figure, the FGM-N method does not produce $Z$ and $h$ variances near the flame front because it neglects the PD effect.
		The FGM-PFN method successfully reproduces the $Z$ and $h$ distributions in the detailed calculation.
		Furthermore, the FGM-PFN method generally agrees with the detailed calculation with respect to the enthalpy decrease near the wall and the $u$ and $T$ distributions.
		However, a slight difference is observed between the flame shapes obtained using each method.
		The flame height of the FGM-PFN method is shorter than that of the detailed calculation and higher than that of the FGM-N method.
		In other words, the slope of the flame front differs between the three methods.
		This difference is also observed in the two-dimensional simulation in the previous section.
		These results suggest that the FGM-PFN method improves the reproducibility of the flame curvature compared with the FGM-N method, but still cannot capture it perfectly.

		Thus, the proposed FGM-PFN method improves the prediction accuracy of the $u$ and $\dot{\omega}_C$ distributions and the flashback speed, while there still exists a discrepancy from the detailed calculation.
		In general, the flame stretch rate $\varepsilon$ is expressed as:
		\begin{equation} \label{eq:stretch2}
			\varepsilon = - \bm{n n} \colon \nabla \bm{u}_f + \nabla \cdot \bm{u}_f + S_d \left(\nabla \cdot \bm{n} \right),
		\end{equation}
		where $\bm{n}$ denotes the vector normal to the flame surface, $\bm{u}_f$ is the velocity at the flame surface, and $S_d$ is the displacement speed.
		The first and second terms in Eq.~(\ref{eq:stretch2}) indicate the flame stretch induced by the velocity gradient tangential to the flame front, whereas the third term indicates the flame stretch induced by the curvature.
		In the FGM-PFN method used in this study, the flamelet database is generated from flames with stretch caused by the velocity gradient.
		In contrast, the stretch of the flashback flames is mainly caused by the curvature.
		This difference in flame properties between the flame in the database and the flashback flame may result in a discrepancy from the detailed calculation.
		\clearpage
	
	\section{Conclusions}
	\label{sec_conclusion}
	In this study, an FGM method that considers the preferential diffusion (PD), flame stretch (FS), and non-adiabatic (NA) effects (FGM-PFN method) was proposed and validated. 
	To investigate the applicability of the FGM-PFN method to the numerical simulations of \ce{H2} flames, two-dimensional numerical simulations of premixed \ce{H2} flame laminar boundary layer flashback were performed using a detailed calculation, and the FGM-PFN, FGM-PN, and FGM-N methods under an ambient pressure of 1 atm, an unburnt temperature of 750 K, and an equivalence ratio of 0.5.
	A three-dimensional numerical simulation of turbulent boundary layer flashback was performed using the detailed calculation, and the FGM-PFN and FGM-N methods under the same conditions.
	
	It was found from the two-dimensional numerical simulations that the FGM-PN method successfully predicted the decrease in the mixture fraction $Z$ near the flame front caused by PD, whereas a constant $Z$ field was obtained using the FGM-N method.
	However, the FGM-PN method failed to predict the variation in $Z$ along the flame surface. 
	The FGM-PFN method successfully reproduced the $Z$ distribution along the flame surface and improved the prediction accuracy of the flashback speed.
	
	In the three-dimensional numerical simulation, the FGM-PFN successfully reproduced the backflow region, which was also observed in previous experiments and DNSs, with lower computational costs.
	Furthermore, by comparing the FGM-PFN and FGM-N methods, it was observed that the PD and FS affect the size of the backflow region, flashback speed, and reaction rate.
	Thus, it is strongly expected that large-eddy simulation (LES) using the FGM method developed herein will be a strong tool for designing new stable \ce{H2}-fueled combustors and for considering optimal operations, in terms of computational accuracy and costs.
	
	\section*{Acknowledgements}
	This work was partially supported by JSPS KAKENHI (Grant Number 22H00192), and by MEXT as “ Program for Promoting Researches on the Supercomputer Fugaku ” (Development of the Smart design system on the supercomputer “ Fugaku ” in the era of Society 5.0)(JPMXP1020210316). This research used computational resources of supercomputer Fugaku provided by the RIKEN Center for Computational Science (Project ID: hp230039).

	\bibliography{mybibfile}
	
	\clearpage

\end{document}